\documentclass[floatfix,reprint, amsmath,amssymb,aps,superscriptaddress,nofootinbib]{revtex4-2}

\usepackage{graphicx}
\usepackage{dcolumn}
\usepackage{bm}
\usepackage{graphicx}
\usepackage[utf8]{inputenc}
\usepackage[export]{adjustbox}
\usepackage{wrapfig}
\usepackage{amsmath}

\usepackage{amsfonts} 
\usepackage{hyperref}
\usepackage{braket}
\usepackage{xcolor}

\newcommand{\bp}{\boldsymbol{\epsilon}}

\begin{document}

\preprint{APS/123-QED}

\title{Segmented Composite Design of Robust Single-Qubit Quantum Gates}

\author{Ido Kaplan $^*$}%
\affiliation{
School of Electrical Engineering, the Iby and Aladar Fleischman Faculty of Engineering, Tel-Aviv University, Tel-Aviv 6997801, Israel.
 }

\author{Muhammad Erew $^*$}%
\affiliation{
 Raymond and Beverly Sackler School of Physics and Astronomy, Tel-Aviv University, Tel-Aviv 6997801, Israel.
 }
 
\author{Yonatan Piasetzky}%
\affiliation{
 Raymond and Beverly Sackler School of Physics and Astronomy, Tel-Aviv University, Tel-Aviv 6997801, Israel.
 }
 
\author{Moshe Goldstein}%
\affiliation{
 Raymond and Beverly Sackler School of Physics and Astronomy, Tel-Aviv University, Tel-Aviv 6997801, Israel.
 }
 
\author{Yaron Oz}%
\affiliation{
 Raymond and Beverly Sackler School of Physics and Astronomy, Tel-Aviv University, Tel-Aviv 6997801, Israel.
 }

\author{Haim Suchowski}%
\affiliation{
 Raymond and Beverly Sackler School of Physics and Astronomy, Tel-Aviv University, Tel-Aviv 6997801, Israel.
 }

\date{\today}

\begin{abstract}

Over the past few decades, quantum information processing research has focused heavily on error mitigation schemes and error-correcting codes. However, while many proposed schemes have been successful in mitigating errors, most of them are perturbative and assume deterministic systematic errors, leaving studies of the problem considering the full noise and errors distribution scarce.
In this work, we introduce an error mitigation scheme for robust single-qubit unitary gates based on composite segmented design that accounts for the full distribution of the physical noise and errors in the system. We provide two optimization approaches to construct these robust segmented gates: perturbative and non-perturbative, which address all orders of errors.
We demonstrate the effectiveness of our scheme in the photonics realm for the dual-rail directional couplers realization. Specifically, we show that the 3-segmented composite design for the fundamental single-qubits unitary operations reduces the error by an order of magnitude for a realistic distribution of errors. Moreover, we demonstrate that the two approaches are compatible for small errors, and significantly reduce the overhead of modern error correction codes.
Our methods are rather general and can be applied to other realizations of quantum information processing units.

\end{abstract}

\maketitle

\def\thefootnote{*}\footnotetext{These authors contributed equally to this work.}

\section{Introduction}

The potential exponential speedup for solving hard computational problems and the possible real-time capability to decrypt classical encryption protocols are the driving forces behind the tremendous research effort invested in quantum information processing (QIP) and quantum computing \cite{40years, Aspuru-Guzik2012, Wendin_2017, doi:10.1126/sciadv.1601540}. Over the last several decades, major theoretical breakthroughs have been achieved, developing quantum algorithms with applications in a variety of problems and fields, including algorithms for combinatorial optimization, quantum machine learning, decryption protocols, and variational quantum algorithms to find the ground state energy of Hamiltonian systems such as molecules \cite{Montanaro2016, Cerezo2021}. Yet, the realization of a quantum information processor is still far away. The major obstacles lie in the inherent systematic errors and stochastic noise of the physical building blocks, which influence state preparation through the measurement process or the unitary operations (gates), the basic ingredients of any quantum algorithm.

The problem of errors and noise is usually dealt with by error mitigation schemes or error-correcting codes. In the former, one attempts to reduce the error using various algorithmic schemes, typically with a small overhead \cite{Gottesman1997, Bravyi1998, Dennis2002, Raussendorf2006, Raussendorf2007a, Raussendorf2007b, Fowler2009, DiVincenzo2009, Fowler2011, Wootton2012, Fowler2013, Vijay2015, Bravyi2018}. In the latter, one constructs logical qubits or quantum gates using many physical qubits, with redundancy and significant overhead that ensures that the logical qubit significantly outperforms the physical qubit \cite{QuantumErrorCorrection}. Most relevant error-correcting codes are stabilizer codes \cite{Gottesman1997}, a prime example being the surface code, having relative tolerance to local errors \cite{Bravyi1998, Dennis2002}. Yet, the capability of fault-tolerant quantum computation of the surface code is conditioned: the probability of errors has to be under certain thresholds for each operation, e.g., single-qubit gates or double-qubit gates \cite{Gottesman1997, Bravyi1998, Dennis2002, Raussendorf2006, Raussendorf2007a, Raussendorf2007b, Fowler2009, DiVincenzo2009, Fowler2011, Wootton2012, Fowler2013, Vijay2015, Bravyi2018}. High-fidelity physical gates are thus extremely important for realizing a useful error-correcting code. An important step towards fault-tolerant quantum computation is to increase the fidelity of single quantum operations, the single unitary gates, which are fundamental building blocks of QIP. This is challenging in the experimental realizations of QIP, where the slightest fabrication defects or an inaccurate coupling strength can lead to errors that include deviations from target driving amplitudes and frequencies.

In recent years, several studies have proposed schemes to enhance the robustness of state-to-state processes \cite{Levitt1979,Shaka1987,Levitt1986,Timoney2008,trapped1,trapped2,atomic1,atomic2,atomic3,Erlich2019,Kyoseva2019} and to devise robust unitary gates in various realizations of quantum information processing (QIP) \cite{PhysRevResearch.2.043194,PhysRevA.101.012321,PhysRevB.102.075311,PhysRevA.104.012609,PhysRevA.103.052612,Torosov:2022gtb,Torosov:2022cdc,9774914}. One of the leading concepts in robust designs is based on the principles of composite pulse sequences used in atomic physics and nuclear magnetic resonance. These sequences use a combination of constant pulses to minimize errors during the evolution of quantum systems \cite{Levitt1979,Shaka1985,Shaka1987,Levitt1986,Timoney2008}. These techniques utilize a perturbative expansion of the gate's operation in small deterministic systematic errors and mitigate the errors order by order. Typically, these schemes deal with varying one parameter of the Hamiltonian. Another variant of this framework for error mitigation is control theory and optimal control. In this approach, an optimal path in the control parameter space is specified to construct a required quantum state from a given initial state or an approximation in the norm sense to a required quantum gate \cite{optimalcontrol1,optimalcontrol2,optimalcontrol3,optimalcontrol4,optimalcontrol5}. However, the control parameters are usually treated as deterministic, or have only one stochastic parameter.

Recently, an expansion of the technique of composite pulses has been proposed to include the full parameter space. Specifically, in integrated photonic-based QIP, which utilizes photons as low-noise carriers of quantum information in the dual-rail representation, fabrication may cause geometrical errors that primarily influence the Hamiltonian's diagonal part. A recent proposed robust solution for state-to-state directional couplers based on composite segmented couplers of different widths \cite{Kyoseva2019} was experimentally demonstrated \cite{9774914}. The design approach showed that modifications to the fabrication protocols are not required.

However, all these proposed composite schemes deal with deterministic errors and noise, whereas, in reality, noise is random by its nature, with randomness inherited from the quantum world, thermodynamic fluctuations, and from errors in manufacturing, preparation, and measurement. These issues become even more acute when dealing with a realization of robust unitary gate operations needed to allow full operation and control of quantum information processors, with a high-enough accuracy to comply with a specific target design for each physical realization. For photonic based realization, for example, this target accuracy is the fourth decimal point \cite{Matthews2009,OIDA:,Pelucchi2022,Moody_2022}, a target that places stringent fabrication tolerances on process parameters such as etching depth, wave-guide widths, etc., which are challenging to meet in practice. While the current known perturbative schemes have succeeded to construct robust gates for the realization of robust unitary gate operations, treatment of the statistical nature of noise and errors is still lacking.

Here we present a scheme for robust unitary operations for realistic quantum platforms. In contrast with previous demonstrations of robust unitary gate designs, we provide protocols that consider the statistical nature of noise and errors in physical systems, and \emph{all} orders of jointly distributed random errors. In devising our robust unitary gates, we follow two design paths. The first one is based on a perturbative approach, where we reduce the fully correlated error order by order in perturbation theory. The second is a non-perturbative method, where we search for the local maxima of the fidelity cost function so that we optimize while accounting for all orders of errors or their variances simultaneously. In order to show the great applicability of our framework, we apply both methods to the photonic dual-rail realization, providing robust high-fidelity unitary solutions to different single-qubit gates, including the fundamental $X$, ${X}^{\frac{1}{2}}$, ${X}^{\frac{1}{3}}$ and Hadamard gates. We demonstrate that the unitary segmented solutions are effective and compatible in practical scenarios of directional-couplers realizations, and are far more robust to systematic errors as compared to uniform couplers. 
Furthermore, we present the advantage of utilizing optimized segmented couplers in reducing the the logical error of the logical state of surface codes by order of  magnitude. Moreover, we show that incorporating these gates in a quantum circuit, such as the QFT algorithm circuit, increases the robustness of entire circuits. This presents a significant advancement over previous works that only focus on correcting errors in individual gates. Our error mitigation techniques have practical applications that can improve the performance of complex quantum algorithms. While we take the integrated photonics path-encoded qubits realization as an example to illustrate the strengths of the scheme on-chip building blocks for quantum applications, the method is rather general and can be applied to any other realization of a QIP device.

Our paper is organized as follows: In Section~\ref{sec: The Method}, we present the single qubit gates and our methods for designing robust ones for a general statistical error model, and illustrate them for an example error model. In Section~\ref{sec: Photonics}, we describe how single qubit gates are physically realized in integrated photonics, describe the error model in the integrated photonics realization, and, using our methods, find and design several robust gates according to a statistical model of fabrication errors in the manufacturing process. We further show how the logical error in a surface code, as a consequence, would behave given our solutions and an error model. In Section~\ref{sec: Discussion}, we summarize and discuss our results. In the appendixes, we provide details of the calculations and further information on various solutions.

\section{Method and Illustration on a Reduced Error Model \label{sec: The Method}}

\subsection{Single Qubit Quantum Gates and Fidelity}

The time evolution of a general qubit system \{$\ket{0},\ket{1}$\} is governed by the Schrödinger equation:
\begin{equation}
    i \partial_t \left( \begin{array}{c}
         c_1\left(t\right)\\
         c_2\left(t\right)
    \end{array} \right)=\begin{pmatrix}
-\Delta\left(t\right) & \Omega^*\left(t\right)\\
\Omega\left(t\right) & \Delta\left(t\right)
\end{pmatrix} \left( \begin{array}{c}
         c_1\left(t\right)\\
         c_2\left(t\right)
    \end{array} \right) ,
    \label{eq: Schrödinger}
\end{equation}
where $c_1\left(t\right)$ and $c_2\left(t\right)$ are the probability amplitudes at time $t$ of the states $\ket{0}$ and $\ket{1}$ respectively, $\Omega\left(t\right)$ is the (complex) Rabi frequency, $\Delta\left(t\right)$ is the (real) detuning, and we set $\hbar=1$. The unitary propagator of such a system is:
\begin{equation}
U \left( t,0 \right)=\mathcal{T}\left[\exp{\left[-i \int_0^t \begin{pmatrix}
-\Delta\left(t^\prime\right) & \Omega^*\left(t^\prime\right)\\
\Omega\left(t^\prime\right) & \Delta\left(t^\prime\right)
\end{pmatrix}\right]} \,dt^\prime \right] .
\label{eq: General Propagator}
\end{equation}
When $\Omega$ and $\Delta$ are independent of time, the propagator simplifies to:
\begin{equation}
 U \left( t,0 \right) =\exp{\left[-i t \begin{pmatrix}
    -\Delta & \Omega^*\\
    \Omega & \Delta
    \end{pmatrix}\right]} .
    \label{eq: Propagator}
\end{equation}

Using physical systems that follow such $\mathrm{SU}\left(2\right)$ dynamics, one can implement various single qubit gates. However, when one considers  noise in the physical system, the implemented gate deviates from the desired one. In order to quantify how far the noisy gate is from the desired one, we will consider the metric provided by the fidelity $F$ of the gate $U(\bp)$, which is defined as
\begin{equation}
F(U_\mathrm{ideal},U(\bp))= \frac{1}{2} |\mathrm{Tr}(U_\mathrm{ideal}^{\dagger}U(\bp)| ,
\label{GF}
\end{equation}
where $U_\mathrm{ideal}$ is the desired ideal unitary gate given by Eq. (\ref{eq: Propagator}), and $U(\boldsymbol{\epsilon})$ is its actual physical realization, which depends on a set of jointly distributed random errors, $\bp= \{\epsilon^a\}_{a=1}^m$. This fidelity takes values in the interval $[0,1]$, where $1$ corresponds to the case of no errors, and $0$ corresponds to the case of maximal deviation from the desired unitary gate operation.

The goal in this work is to increase the expectation value
of the fidelity: 
\begin{equation}
    \bar{F} = \mathbb{E}_{\bp}[F(U_\mathrm{ideal},U(\bp)] \ .
    \label{AF}
\end{equation}
This will be done both for the general case as well as for a specific statistical error model of integerated photonic realm. The relevant statistical error model should be taken depending on the specific physical realization of the gates, where one considers the quantum errors, thermodynamic errors, and the errors of manufacturing, preparation, and measurement. 
Maximizing the mean fidelity over a wide error range is crucial for fault-tolerant computation, as mentioned in the introduction, since a certain  threshold for the resulting physical error probability has to be achieved.


\subsection{Constructing Robust Composite Gates}

The method that we employ to design robust gates is to compose pulses or segments. The reasoning
behind this approach is the natural assumption that the relevant errors are highly correlated, and this correlation can be applied to cancel errors with appropriately tuned designs.
Consider an ideal unitary gate $U_\mathrm{ideal}$, as well as its actual noisy segmented realization $U^{(N)}=\prod_{k=1}^N U_k(\bp_k)$, where $\bp_k$ is the random error vector of the $k^{th}$ segment, which includes $m$ errors: $\bp_k=\{ \epsilon_k^a \}_{a=1}^m$. All the errors are jointly distributed random variables. Each segment $U_k$ without errors is as in Eq. (\ref{eq: Propagator}).
The goal is to increase the expectation value
of the fidelity in Eq. (\ref{AF}).

In our analysis, we employ two methods. The first one is perturbative in the error random variables, where we consider
them to be fully correlated and design the segmented gate such that we cancel the errors
order by order in perturbation theory. More specifically, we construct analytical solutions
of 3-segmented
designs that cancel the first order error term. Clearly, cancellation of higher order error terms requires
a larger number of segments.

The second method is non-perturbative, where we consider Eq. (\ref{AF})
as a cost function to be maximized. 
While these two methods are compatible for small errors or small variances of errors, as will be seen, the non-perturbative approach also offer a path for addressing large values of the random error variances, where the optimization take into account all orders in the errors simultaneously.


 
\subsection{Example: A detuning Error Model \label{sec: detuning error model}}


In order to illustrate our methods in a relatively simple case, we consider first a physical system which allows only real $\Omega$'s, and we assume a single error random variable $\epsilon_k = \epsilon, k=1,...,N$,
which is a systematic error in $\Delta$, and neglect the error in $\Omega$. We further assume that the errors of the different segments are 
fully correlated.
This assumption describes well the errors in several quantum and classical systems that follow such  $\mathrm{SU}\left(2\right)$ dynamics, such as gates of trapped ions \cite{trapped1,trapped2}, sum-frequency generation \cite{boyd2020nonlinear}, atomic systems \cite{atomic1,atomic2,atomic3}, etc.

The $N$-segmented gate reads:
\begin{equation}
    U^{\left(N\right)} = \prod_{k=1}^N   U \left(\Omega_k,\Delta_k,t_k,\epsilon\right) \ , 
\end{equation}
where
\begin{equation}
    U \left(\Omega,\Delta,t,\epsilon\right)=e^{-i t\left(\Omega X-\Delta Z-\epsilon Z \right)} \ ,
\end{equation}
and $t_k,\Omega_k,\Delta_k$ are the length of the $k^{th}$-segment, its coupling, and its detuning, respectively.
The $N$-segmented gate fidelity (\ref{GF}) is
$F(U^{(N)}(0),U^{(N)}(\epsilon))$.

\subsubsection{Perturbative Method}
In the perturbative approach, we consider the error in the quantum gate
\begin{equation}
E(\epsilon) = U(\epsilon)-U(0) = \sum_{k>0} E_k \epsilon^k \ .  
\label{E}
\end{equation}
The task is to design an $N$-segmented gate $U$ such that $E = O(\epsilon^n)$ for a given $n$. In many practical realizations, it is sufficient to take $n=2$.
Note that since $E(\epsilon)$ is a function of a random variable instance, removing the linear order
term is not simply removing the expectation value of $\epsilon$.

Let us take for example $U=X$, and construct the gate up to an overall phase, that is, $iX$.
We need
to find 
$N$ and $\{ \Omega_k \}_{k=1}^{N},\{ \Delta_k \}_{k=1}^{N} ,\{ t_k \}_{k=1}^{N}$ such that $U^{\left(N\right)}\left(0\right)=iX$, $ \left. \frac{\partial E^{\left(N\right)}\left(\epsilon\right)}{\partial \epsilon}\right| _{\epsilon=0}=0$ , $ \left. \frac{\partial ^2 E^{\left(N\right)}\left(\epsilon\right)}{\partial \epsilon ^2}\right| _{\epsilon=0}=0$, etc. Using the unitarity of each propagator one can simplify the equations and reduce their complexity.
In Appendix~\ref{appendix: Analytical Solutions for The Toy Model} we present analytical robust solutions of the equations
for the gates of the form ${\left(iX\right)}^{\frac{1}{n}}$, where $n$ is a positive integer, employing three segments, and two solutions of the $iX$ gate using four segments with the same coupling constant.
We also compare in Appendix~\ref{appendix: Analytical Solutions for The Toy Model} our solutions' fidelity to that of a single segment for $n=1,2,3$.

\subsubsection{Non-Perturbative Method}

In the non-perturbative approach, we search for a maximum of a cost function (minimum of a loss function) by optimization. In order to simplify the optimization process, we split the loss function into two subfunctions, the \emph{invalid range loss subfunction} and the \emph{robust fidelity loss subfunction}. The former ensures that the parameters we obtain are within their allowed range (for instance, the length of the waveguides cannot be negative) by strongly penalizing deviations from it.
The robust fidelity loss subfunction calculates the fidelity for a range of $N$ error values between $-3\sigma$ to $3\sigma$, $\sigma$ being the standard deviation, and weighs these fidelities according to the assumed error distribution (for instance, a normal distribution). 
An overall minus sign is added in order for the algorithm to minimize this value and thus maximize the overall fidelity. For example, the loss function used for errors that have a normal distribution is:
\begin{equation}
    \begin{split}
        \mathrm{Loss} = 1 - \sum_{x=-3 \sigma }^{3 \sigma } \frac{e^{-\frac{x^2}{2 \sigma^2}}}{\mathrm{Dist}_\mathrm{sum}}  \cdot F(U_\mathrm{ideal},U_\mathrm{optimization})(x)  \\ + \mu \cdot \sum_{i=0}^{N-1}  \sum_{j=0}^{m-1} \max(0, p_{j,i} - p_{j}^{\mathrm{Max}}) + \max(0, p_{j}^{\mathrm{Min}} - p_{j,i}) \ ,
    \end{split}
\end{equation}

The former sum in the loss function is discrete: between each pair of subsequent $x$ values there's an interval of $\frac{1}{n}$, where $n+1$ is the number of samples used to estimate the integral. The value $\mathrm{Dist}_\mathrm{sum}=\sum_{x=-3 \sigma }^{x=3 \sigma} e^{-\frac{x^2}{2 \sigma^2}}$ is used to normalize the distribution function, which guarantees that the robust fidelity loss subfunction's minimal value is 0.  
$\mu$ defines the weight ratio between the loss subfunctions. There are $N$ segments used, with $m$ parameters per segment (for instance, in this model, $m=3$ because there are three parameters: $ \Omega , \Delta, t$). $p$ is the selected parameter values. $p_{j}^\mathrm{Min}$ and $p_{j}^\mathrm{Max}$ are chosen by physical limitations (for instance, $t^\mathrm{Min}=0$, since the length of the waveguides cannot be negative) and $\mu$ is set to be in the range of $[5,100]$.

By minimizing these two subfunctions, we obtain physically feasible parameters which minimize the fidelity loss for errors between $-3\sigma$ to $3\sigma$ weighted by the given error distribution. Furthermore, the optimizer we used is the Adam  optimizer \cite{kingma2017adam} (Adaptive Moment Estimation optimizer), which is an optimizer that  computes individual adaptive learning rates for different parameters from estimates of first and second moments of the gradients. The initial learning rate we used is $10^{-3}$.

\begin{figure}[htbp]{}
        \includegraphics[width=8.5cm,clip,trim=2.2cm 2cm 2.5cm 1.5cm]{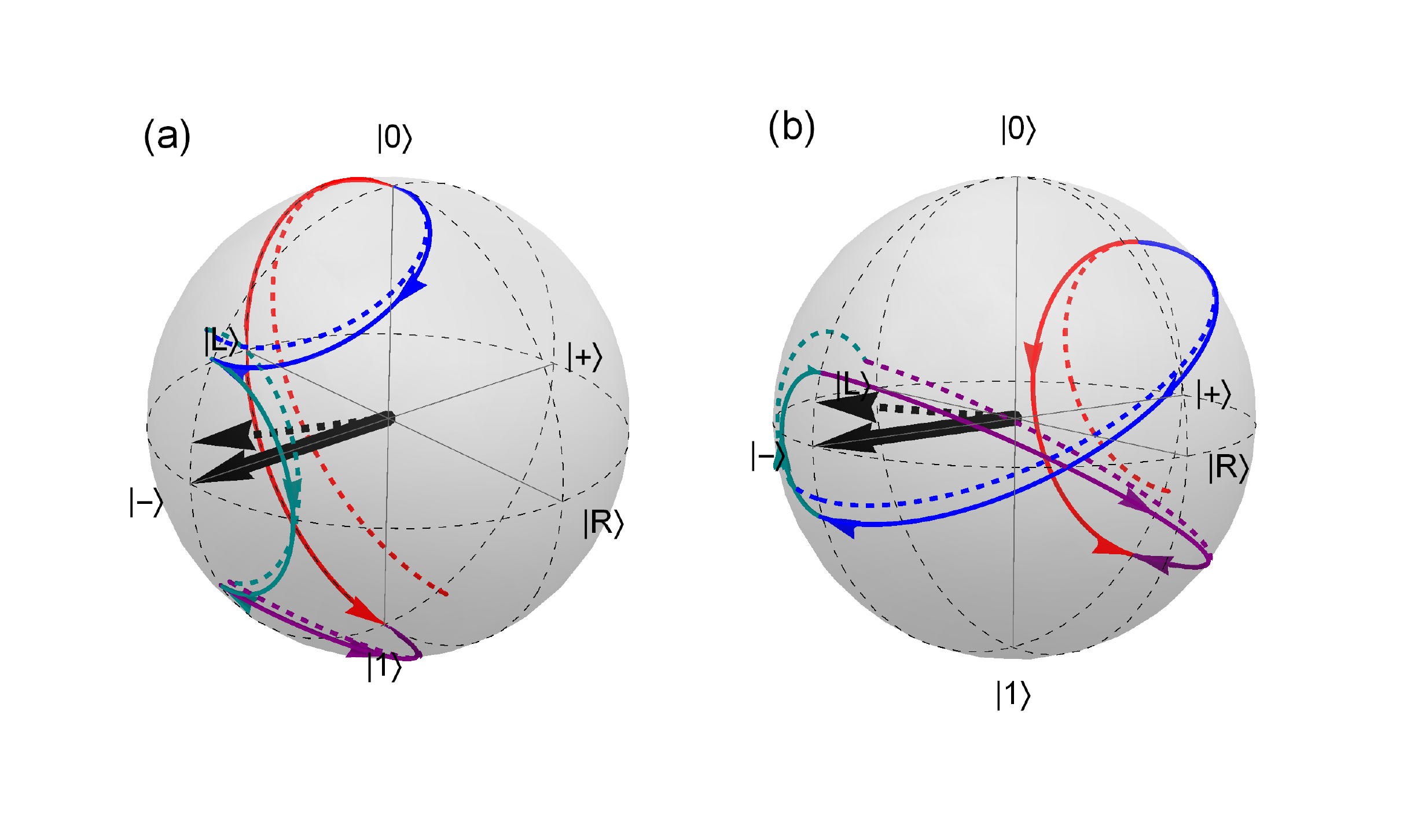}
        \includegraphics[width=0.48\textwidth,page=4,clip,trim=0cm 11cm 0cm 0cm]{FiguresForrobustGates.pdf}
        \caption{Na\"{\i}ve vs. composite gate. (a-b) Bloch sphere representation of a robust composite $-iX$ Gate. The plot provides a schematic description of two different states on Bloch sphere and the trajectories they follow under the uniform $-iX$ gate (in continuous red), and under the segmented gate presented in Table \ref{table: Non-Perturbative examples For Toy Model} (in continuous blue, turquoise, and purple). In dashed lines, we show the trajectories the states follow when an error $\epsilon=0.17\Omega$  occurs simultaneously in all detunings. In black, we depict the torque vector of the uniform gate around which the states rotate under the $-iX$ operation.
        One can see the robustness of the segmented gate against such errors compared to the uniform regular gate. We show this for two different initial states, $\ket{0}$ and $\cos\left( \frac{\pi}{8} \right)\ket{0}+\sin\left( \frac{\pi}{8} \right)\ket{1}$, to emphasize that the whole gate is robust, not only the complete population transfer between $\ket{0}$ and $\ket{1}$. In other words, the $-iX$ gate is robust for any initial state; i.e. not only the magnitude of the element $U_{12}$ of the representative matrix of the operation is robust (against errors in the physical system) but also its phase, as well as the phase of the element $U_{11}$. (c-d) Dual rail photonic realization of unitary gates. Schematic 2D top view illustrations of standard and composite gates respectively, based on the directional couplers realization of gates in integrated photonics. We denote the waveguides widths $w_1,w_2$, the length $t$ and the gap $g$. }
         \label{fig: Bloch}
     \end{figure}

Examples of non-perturbative solutions for the detuning error model and their simulations can be seen in Appendix~\ref{appendix: Non-Perturbative Solutions for The Toy Model}.
We show in Fig.~\ref{fig: Bloch}(a-b) one solution on the Bloch sphere compared to the uniform gate, as well as how errors affect the result of the gate for two different initial states for each case (uniform and composite).

 Further details regarding the optimization process are described in Appendix \ref{appendix: numeric}. 
 
\section{Robust segmented gates in Integrated Photonics \label{sec: Photonics}}

As there are several realizations of quantum gates and each one has an appropriate statistical model of errors, we choose, the following, to apply our methods to the photonic realm \cite{OIDA:,Pelucchi2022,Moody_2022}.
This realm, which utilizes photons as excellent low-noise carriers of quantum information, requires that unitary gate  comply with a target design to a fourth decimal point accuracy\cite{Matthews2009}.

\subsection{Directional Couplers as Gates and their Error Model \label{sec: Error Model for correlated}}

According to the coupled-mode theory, the propagation of the pair of electrical fields $E_{1,2}$ in a directional coupler of a fixed cross-section is described exactly by  Eqs.~(\ref{eq: Schrödinger}) and~(\ref{eq: Propagator}), where the actual matrix elements that describe the dynamics along the two waveguides are the mode propagation constants' mismatch $\Delta\beta$ and the interaction coupling $\kappa$ between the two waveguides \cite{boyd2020nonlinear}. The coupling coefficient $\kappa$ between the waveguides is equivalent to the off-diagonal term $\Omega$. The mode mismatch between the mode index $\Delta\beta=\beta_1-\beta_2$ is equivalent to the diagonal term $\Delta$, and the propagation length $z$, is equivalent to the evolution time $t$.  The  coupling $\kappa$ is largely determined by the distance between the cores.

In our analysis of the functional dependence of the coupling, the detuning, and the relevant error model, we solve for the realization of single-mode silicon-on-insulator rib waveguides. The parameters $\Omega$ and $\Delta$ in Eqs. (\ref{eq: Schrödinger}) and (\ref{eq: Propagator}) are in fact functions of the following physical parameters, some of which are depicted in Fig.~\ref{fig: Bloch}(c),(d):
\begin{enumerate}
    \item $h_1, \: h_2$ --- Etching depths,
    \item $w_1,\: w_2$ --- The widths of the waveguides,
    \item $H_1,\: H_2$ --- The heights of the waveguides,
    \item $g$ --- The gap between the waveguides,
    \item $T$ --- The temperature,
    \item $\lambda$ --- The wavelength.
\end{enumerate}
waveguide devices can support low-loss bends, down to some finite radius, mostly determined by the refraction index contrast between the core and the cladding of the waveguide. Below this radius, significant losses occur due to scattering from wall roughness and radiation loss from the curvature of the waveguide \cite{Bahadori2019}. Typical Silicon on Insulator (SOI) devices usually allow a bend radius to be no smaller than roughly 10 microns. This results in difficulty in applying significant gap changes abruptly (i.e., within a distance that is considerably less than the length of a segment). Thus, in our designs we aimed for a fixed gap for all the different segments. For a fixed gap, etching depth, temperature, wavelength, and waveguide heights:
\begin{subequations}
    \begin{equation}
        \Omega=\kappa\left(w_1,w_2\right) ,
    \end{equation}
\begin{equation}
        \Delta=\Delta\beta\left(w_1,w_2\right) .
\end{equation}
\end{subequations}
In order to estimate these functions, i.e., the mode propagation mismatch and coupling coefficients as functions of the geometric components for a desired range of values, we used the coupled mode theory approximation, Lumerical simulations, and known fitting methods. For details, see Appendix \ref{appendix: Fit}.

We assume that for the desired set of widths of the waveguides, they all have the same error, i.e. they are fully correlated, and this error is distributed normally:
\begin{equation}
\delta w \sim \mathcal{N} \left(0,\sigma^2\right),
\end{equation}
independently of the value of the desired set of widths. We describe our perturbative method in Section \ref{sec: method for rho=1 - purturbative}, and the non-perturbative numerical search in Section \ref{sec: method for rho=1 - non-perturbative}.

\subsection{Methodology\label{sec: method for rho=1}}

Using the interpolation functions for the dependence of $\kappa$ and $\Delta\beta$ on the parameters, and assuming all segments have the same error in widths, the error model can be dealt with perturbatively in a simple way.
We define the $k$th segment of the $N$-segmented gate by
\begin{equation}
 U_k = 
    e^{-i \frac{z_k}{2}\left(\kappa\left({w_1}_k+\delta w,{w_2}_k+\delta w \right) X-\Delta\beta\left({w_1}_k+\delta w,{w_2}_k+\delta w \right) Z \right)} ,
\end{equation}
where $z_k,{w_1}_k,{w_2}_k$ are its length and widths respectively.
The $N$-composite gate reads:
\begin{equation}
U^{\left(N\right)}\left(\delta w\right) = 
U_c \left(\prod_{k=1}^N U_k \right) U_c ,
\label{eq: composing2}
\end{equation}
where the matrix $U_c$ represents the non-zero coupling effect that occurs when the two waveguides are brought closer and taken further away. We model this effect as another segment at the beginning and the end of the composite segment design, with zero detuning,
given by
\begin{equation}
\label{eq:Uc}
 U_c = \cos(\theta_c) I - i \sin(\theta_c)X .
\end{equation}
The parameter $\theta_c =  0.232$ was determined numerically following \cite{EmreKaplan:19}, and verified experimentally by fabricating various zero detuning directional couplers with an identical cross-section, measuring the coupling ratio, extrapolating the coupling ratio to zero coupling length, and finally estimating the amount of coupling that occurs only from initiating and terminating the interaction.

\subsubsection{Perturbative Method \label{sec: method for rho=1 - purturbative}}

We seek solutions that make the derivatives vanish:
\begin{equation}
    \left. \frac{\partial^j U^{\left(N\right)}\left(\delta w\right)}{{\partial \delta w}^j}\right| _{\delta w=0}=0,
\end{equation}
for $j=1,2,3,\cdots$
When we find a solution, we will provide a plot of its fidelity based on this simplified model, and a plot of its fidelity based on the model described in Section \ref{sec: Error Model for correlated}. One can see from Fig.~\ref{fig: mean perturbative and non perturbative solutions} that it is sufficient for our purposes to work with the former one. This is due to the assumption that $\sigma<20$~nm, which is much smaller than $w_1$, $w_2$.

\subsubsection{Non-Perturbative Method \label{sec: method for rho=1 - non-perturbative}}

\begin{figure}[tb]{}
        \includegraphics[width=0.48\textwidth,page=5,clip,trim=0cm 3cm 0cm 0cm]{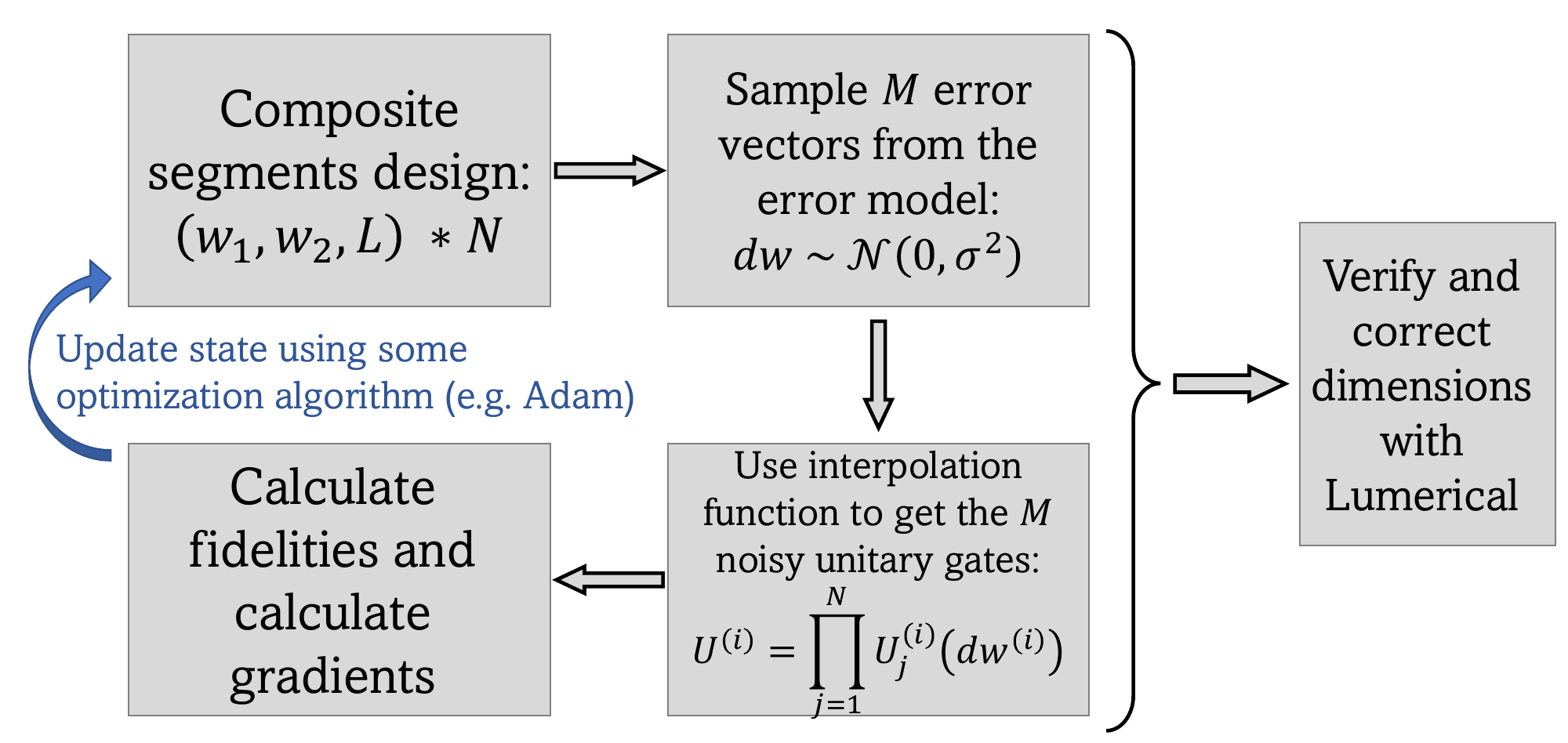}
        \caption{\textbf{The non-perturbative optimization method.} In this method, we initialize our composite design with some randomly chosen geometeric parameters. We then sample geometries from our error model and calculate the resulting fidelities. Using these results, we perform multiple optimization iterations until convergence to a robust design. Lastly, we analyze the resulting design back in Lumerical and fine-tune the lengths.}
         \label{fig: The non-perturbative optimization method}
     \end{figure}

For the numeric approach, we set a correlated error distribution in the waveguide widths, multiply the matrices $M_i$ in stage 4 of the robust fidelity loss calculation described in Appendix~\ref{appendix: numeric} by $U_c$ on both sides to simulate the coupling effect before and after the waveguides enter the directional coupler, and use the interpolation functions in order to translate between the geometric parameters and $\kappa$ and $\Delta\beta$.  We then optimize the geometric parameters of our segmented design, by using a stochastic gradient-based optimization method. Lastly, we analyze and verify the resulting segments in Lumerical. We also correct small discrepancies in the segment lengths that may arise between the coupled mode theory-based coupling approximation (Appendix \ref{appendix: Fit}) and the more accurate two-waveguides simulations with Lumerical. The method for fixing this discrepancy is also explained in Appendix \ref{appendix: Fit}.
The numerical non-perturbative method is illustrated in Fig.~\ref{fig: The non-perturbative optimization method}.

\subsection{Solutions}
In this section, we present selected composite based designs for robust unitary gates in the integerated photonic realm.
The designs, generated both for perturbative and non-perturbative approaches, are compared with the uniform coupler fidelity, which is calculated up to a global phase. The uniform coupler parameters and segmented coupler parameters are fully presented in Appendix \ref{appendix: gate_parameters} in tables~\ref{table: Analytic robust gates against correlated errors in widths} and~\ref{table: Numeric robust gates against correlated errors in widths}. 

\subsubsection{Random error simulations}
\begin{figure}[!h]
     \centering
     \includegraphics[width=8.5cm]{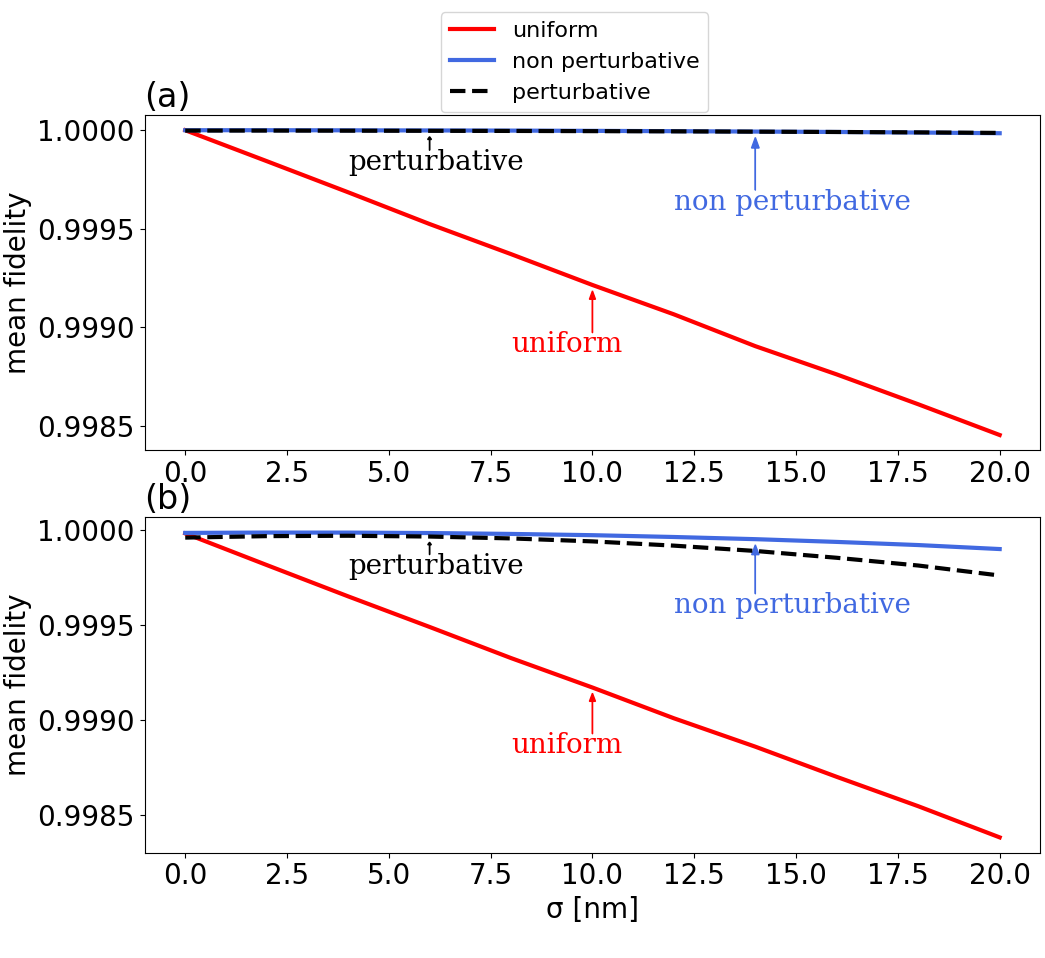}
     \caption{The mean fidelity of the pertubative and non-pertubative composite gates compared to uniform gates, with full error correlation in the width, as a function of the error standard deviation $\sigma$. In (a) the ideal gate is $X$ and in (b) the ideal gate is the Hadamard gate.
   }
     \label{fig: mean perturbative and non perturbative solutions}
\end{figure}

\begin{figure}[!h]
     \centering
     \includegraphics[width=7.5cm]{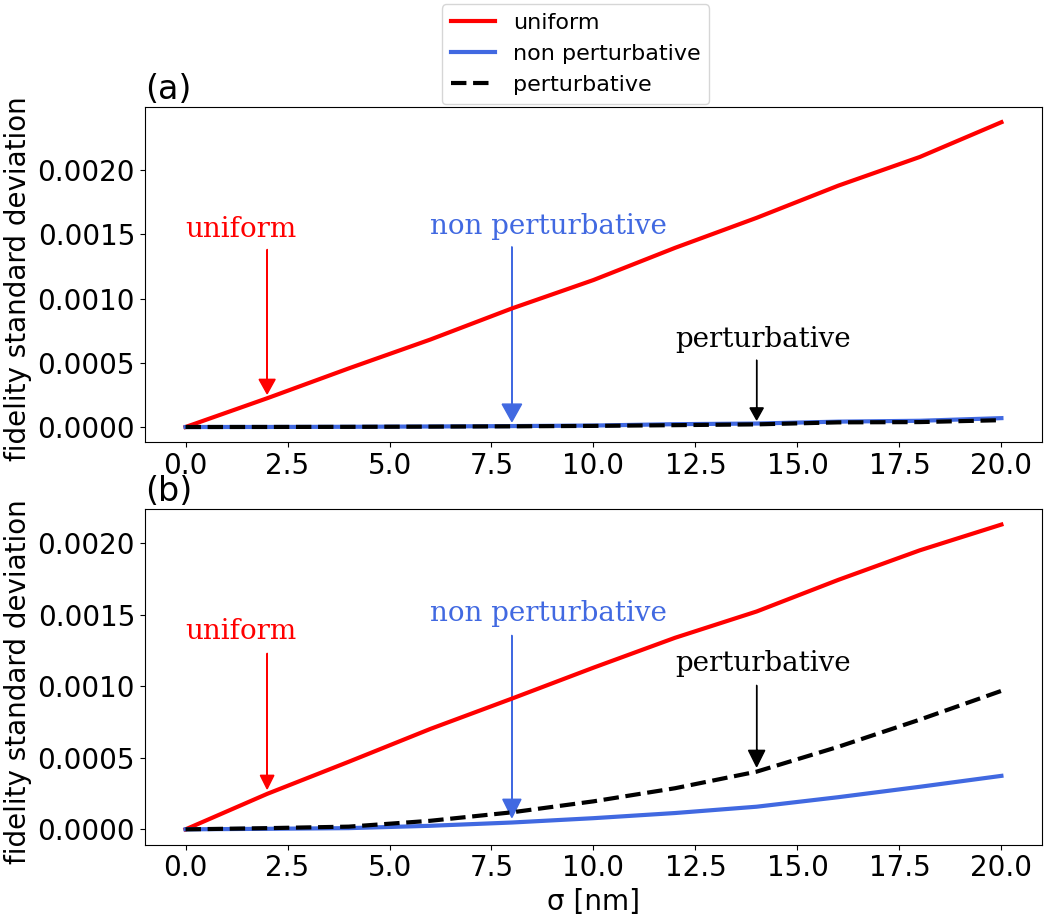}
     \caption{The standard deviation of the fidelity for uniform couplers compared to segmented couplers, with full error correlation in the width, as a function
     of the error standard deviation $\sigma$. In (a) the ideal gate is $X$ and in (b) the ideal gate is the Hadamard gate.}
     \label{fig: fidelity std perturbative and non perturbative solutions}
\end{figure}

In Figs.~\ref{fig: mean perturbative and non perturbative solutions} and~\ref{fig: fidelity std perturbative and non perturbative solutions} we compare the uniform and segmented gates robustness by using width errors sampled randomly from a normal error distribution.
In both simulations, $10^5$ values were sampled for improved accuracy.

In Fig.~\ref{fig: mean perturbative and non perturbative solutions} we compare the mean fidelity of the segmented and uniform gates, assuming fully correlated errors in all widths for all segments. As can be seen from the figure, for both the $X$ gate and the Hadamard gate, there is a clear advantage for the segmented design, which becomes more pronounced as the standard deviation increases.

In Fig.~\ref{fig: fidelity std perturbative and non perturbative solutions} we compare the standard deviation of the fidelity of the segmented and uniform gates, assuming again fully correlated errors in all widths for all segments. We see that for every $\sigma$ value in the given range, the  standard deviation of the fidelity of the segmented design is lower than that of the uniform one. Furthermore, we can see that for the $X$ gate simulation (Fig.~\ref{fig: fidelity std perturbative and non perturbative solutions} (a)), as the  value of $\sigma$ increases, the difference in fidelity standard deviation between the segmented and uniform design rises linearly. This means that compared to the uniform design, the segmented design is far less likely to suffer from random fidelity values lower than the mean fidelity, even when the average width error in the waveguides is greater.

\subsubsection{Deterministic error simulations}

\begin{figure}[!h]
     \centering
     \includegraphics[width=8cm]{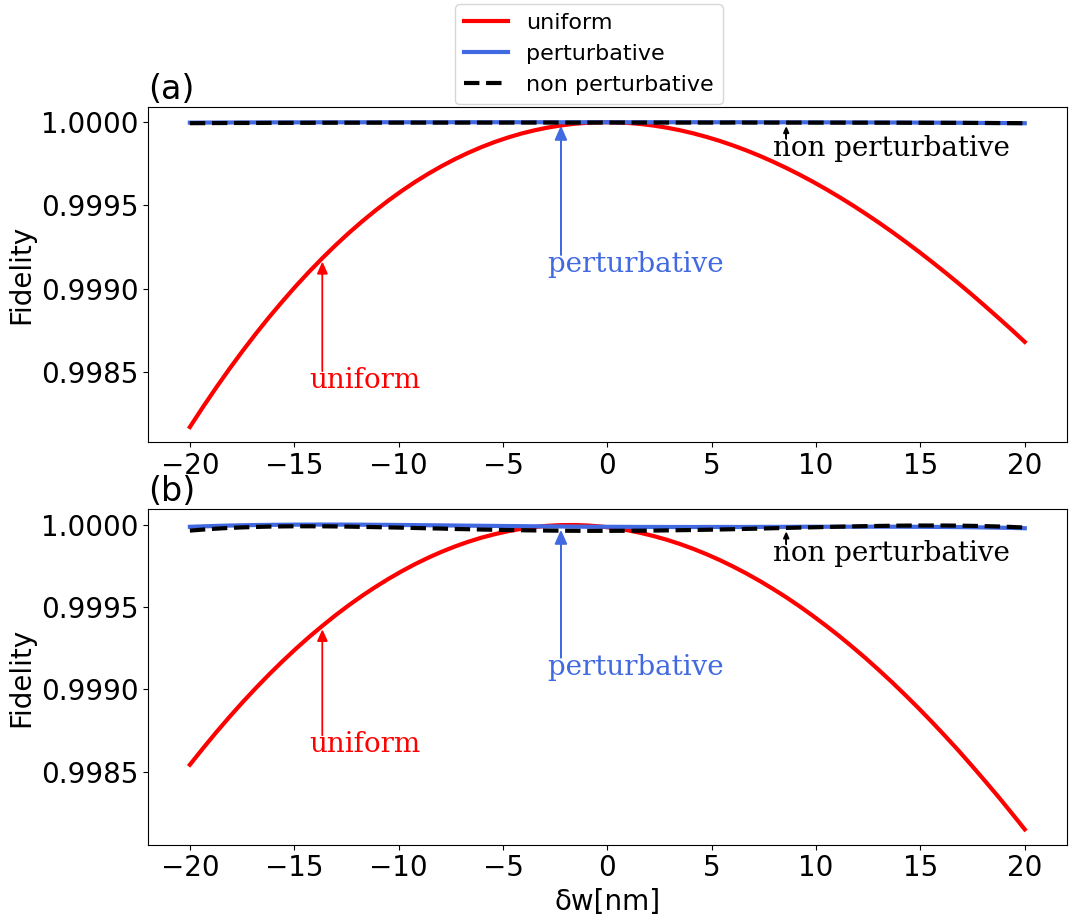}
     \caption{The fidelity of the pertubative and non-pertubative composite gates compared to uniform gates for a fixed error value in the waveguide widths, $\delta w$. In (a) the ideal gate is $X$ and in (b) the ideal gate is the Hadamard gate.}
     \label{fig: deterministic perturbative and non perturbative solutions}
\end{figure}

In Fig.~\ref{fig: deterministic perturbative and non perturbative solutions} we compare the uniform and segmented gates robustness for fixed deterministic errors. In the simulations, the fidelity of both uniform and segmented couplers was calculated for multiple error values between -20~nm and 20~nm, that is between -3 and 3 standard deviations.
As shown for the random error simulations, the fidelity of the segmented design is far more robust in comparison to the uniform one.
Furthermore, for both the perturbative and non-perturbative approaches, the difference between the fidelity of the uniform and segmented couplers increases in a parabolic fashion as $\sigma$ increases.

As seen clearly in the figures in this section, the segmented designs are more robust than the uniform ones, having higher fidelity mean and lower fidelity variance.

\subsection{Logical Error}

The error reduction shown in Fig.~\ref{fig: mean perturbative and non perturbative solutions} and \ref{fig: deterministic perturbative and non perturbative solutions} demonstrates the mitigation 
of correlated physical errors, and is evidently important during the Noisy Intermediate-Scale Quantum quantum computation (NISQ) era \cite{NISQ}, in which a significant error reduction allows an order of magnitude increase in the number of operations one could perform before the circuit becomes too noisy. Moreover, and even much more crucial, error mitigation is also of much relevance to fault-tolerant quantum computers,
where a quantum error-correcting code is implemented. Consider for instance the surface code (for a review see \cite{surfaceCodes}).
The logical error $P_L$ is related to the physical error $p$ by the empirical formula:
\begin{equation}
    P_L \sim (p/p_\mathrm{th})^{d_e} \ ,
\label{surfaceCodeEquation}
\end{equation}
where $p_\mathrm{th}$ is the surface code threshold and is estimated as $0.57 \%$, $d$ is the size of the surface array, and $d_e=(d+1)/2$ is the code distance.
Using equation \ref{surfaceCodeEquation}, we can estimate how close are the uniform and segmented couplers' error rate 
to the empirical surface code threshold for single-qubit gates.

\begin{figure}[!h]
     \centering
     \includegraphics[width=8.5cm]{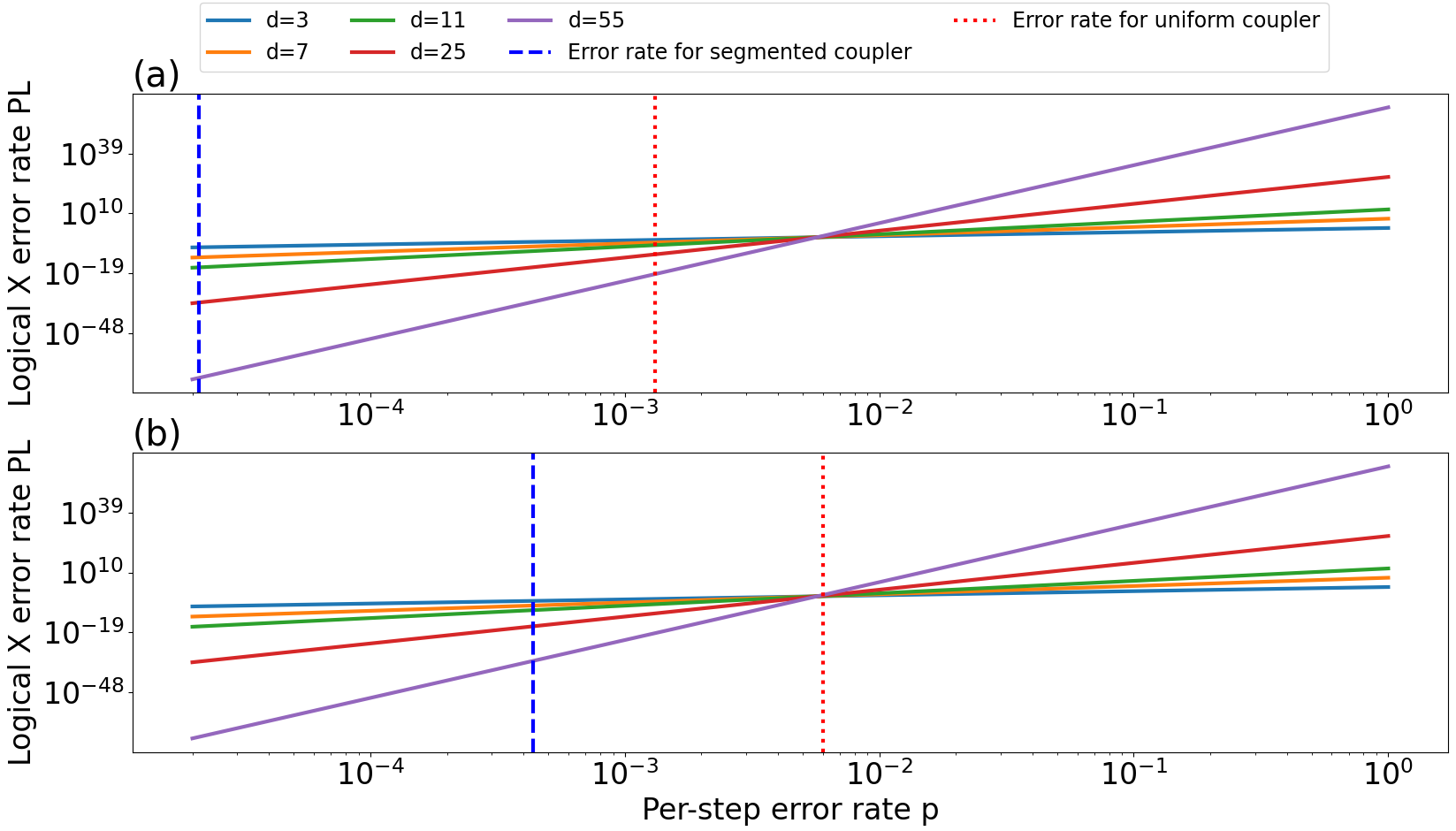}
     \caption{The logical error rate $P_L$ as a function of the physical error rate $p$, tested on X gate using a surface code. In (a) the width error is set to be 3 standard deviations while in (b) the width error is set to be $\sim 15 \%$ of the nominal width (roughly 9 standard deviations). The dashed vertical line denotes the threshold of the quantum error correcting code. 
     }
     \label{fig: Logical Error Rate}
\end{figure}


To estimate the physical error rate $p$, we employed a large number of single logical qubit states $\psi$. For each error rate estimation, we generated 10,000 uniformly random $\psi$ states, where each $\psi$ is defined as
\begin{equation}
    \ket{\psi} = \frac{\ket{\psi_r}}{\sqrt{\braket{\psi_r|\psi_r}}},
\end{equation}
with $\psi_r = (a_r + i a_i)\ket{0} + (b_r + i b_i)\ket{1}$. Here, $a_r$, $a_i$, $b_r$, and $b_i$ are sampled from a uniform distribution between 0 and 1.
Subsequently, we computed the physical error rate for each $\psi$ using the formula
\begin{equation}
    p_{\psi} = \left| \langle U \psi, U_{\mathrm{ideal}} \psi \rangle \right|.
\end{equation}
Finally, our physical error rate $p$ was determined as the minimum value within this range:
\begin{equation}
    p = \min_{\psi} (p_{\psi}).
\end{equation}

The results of this numerical estimate can be seen in Fig.~\ref{fig: Logical Error Rate}, where the parameters used for this optimization are given in table \ref{table: Numeric robust gates against correlated errors in widths}. 
In Fig.~\ref{fig: Logical Error Rate} (a) the physical error rates for uniform and segmented couplers are smaller than the threshold, meaning the logical error in both couplers can be reduced by using the surface code error correction. However, the ratio between the logical error rates is significant and rises exponentially for increasing $d$ values. This implies that, by using the segmented coupler, one can perform error correction efficiently and with less resources (fewerphysical qubits and quantum gates).
In Fig.~\ref{fig: Logical Error Rate} (b), while the physical error rate for the segmented coupler is still smaller than the threshold, the physical error rate for the uniform coupler is not. This means that errors generated in the uniform coupler cannot be corrected using the 
error-correcting surface code. Note that this result is obtained when the width error is set to be very large (above 8 standard deviations). 
In these simulations, we used a circuit model-based quantum error correction code and not a measurement-based quantum computation model \cite{Raussendorf2003}. While this can lead to inaccuracies, since the optimization model itself is applicable to other qubit implementations (whereinstead of segmented couplers, we can use, e.g., composite pulses), we expect these results to be qualitatively correct for photonic systems.

\subsection{Quantum Fourier Transform (QFT) infidelity estimation }

Based on our composite unitary gates, here we show the great improvement in the realization a real quantum algorithm. We choose the QFT, which is basic and critical quantum algorithm in many state-of-the-art quantum algorithms, such as Shor's algorithm. 

\begin{figure}[!h]
     \centering
     \includegraphics[width=8.5cm]{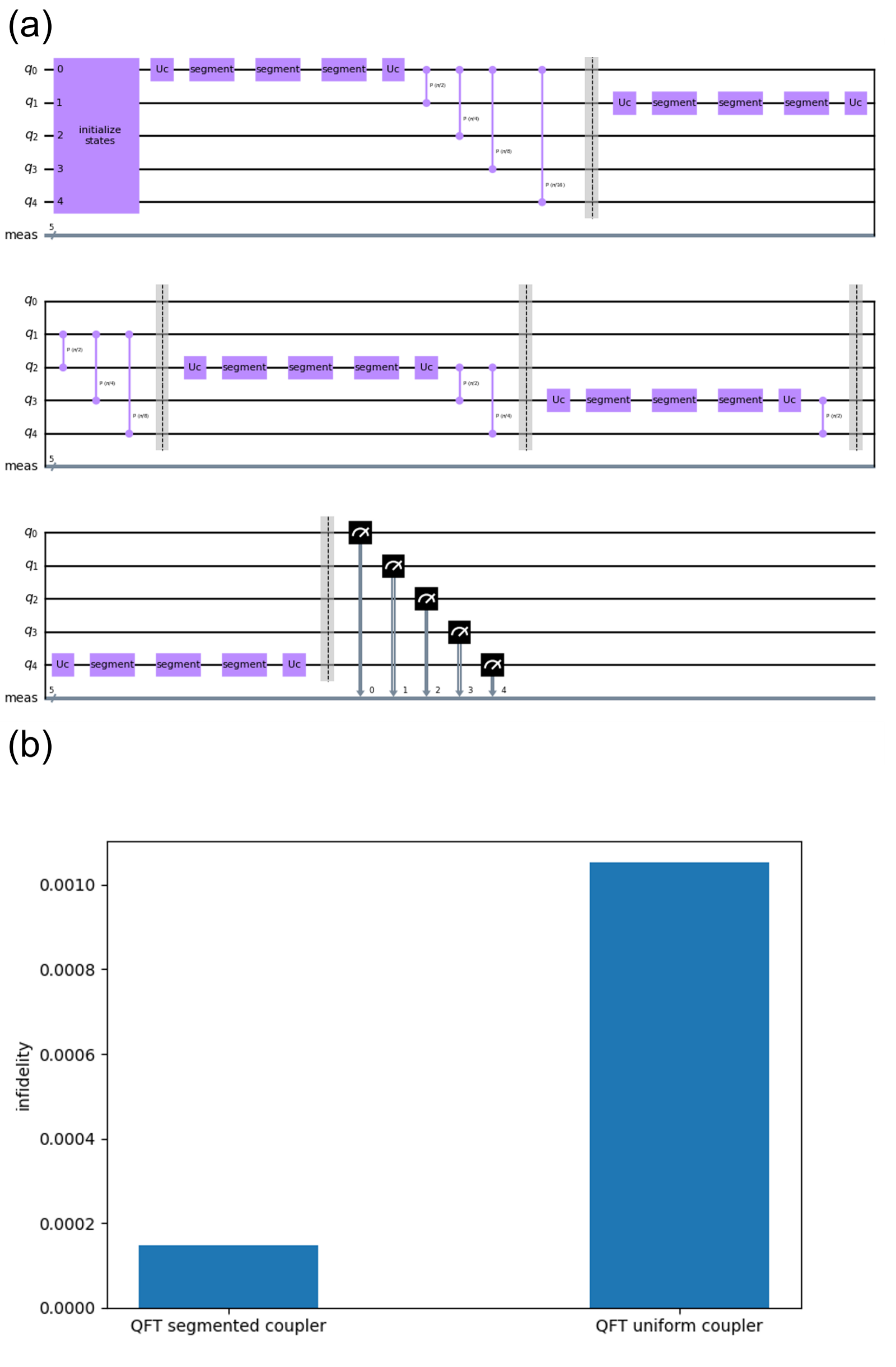}
     \caption{Five qubit QFT algorithm. (a) A circuit-based schematic of a Five qubit QFT algorithm. Each of the gates in the algorithm is chosen to be either traditional-uniform gate or composite-based unitary gates. (b) The infidelity of the QFT cirucit using segmented and uniform coupler}
     \label{fig: QFT circuit}
\end{figure}

In Figure \ref{fig: QFT circuit} (a), we present a model which enables the estimation of the QFT circuit infidelity. We first replace all single qubit gates with directional couplers with given waveguide width error. Afterwards, we initialize the circuit with state: $QFT_{inv}|0_00_1...0_{n-1}>$. Finally, we run the QFT algorithm with the noisy gates and measure the outcome. If the width error is 0, then the algorithm's output will be state $|0_00_1...0_{n-1}>$.
Using this notion, we calculate the circuit's  infidelity in the following manner:
$INF_{QFT}=<0_00_1...0_{n-1}| \overline{QFT} \cdot QFT_{inv}|0_00_1...0_{n-1}>$, where $\overline{QFT}$ is the $QFT$ circuit with noisy single qubit gates.

Using this method, we numerically calculated $INF_{QFT}$ for both the segmented and uniform directional couplers by measuring the algorithm outcomes for 1,000,000 simulations. 

Employing the paramters shown in table \ref{table: Numeric robust gates against correlated errors in widths}, we estimated that the infidelity of the QFT cirucit using the uniform couplers is 6.5 times greater than the infidelity of QFT cirucit using the segmented couplers, as can be seen in Figure \ref{fig: QFT circuit} (b).

The QFT model and our numerical verifications are available in the following Github repository:
\url{https://github.com/Ido-Kaplan/QFT_error_correction}.

\newpage

\section{Discussion \label{sec: Discussion}}

In our research, we considered the fidelity function as a random variable which inherits its randomness by being a function that depends on fully correlated jointly distributed errors. 

These errors are the systematic coherent errors due to imperfections in the construction of quantum devices. Our framework studies the mean and variance of the fidelity function with respect to the joint probability distribution function and optimizes them. The question that we address is whether segmented composite pulses allow for better fidelity properties as specified in this framework. 

While our scheme combines ingredients of the composite pulse and optimal control frameworks, such as gradient ascent algorithm \cite{KHANEJA2005296, chen2022iterative} and trajectory optimization \cite{propson2021robust}, it is not identical to either of them. 

One could have attempted to make a mathematical mapping between the different schemes, but this would require considering the control parameters as a jointly distributed random variables with a particular auto-correlation at different discrete times. This can make the control theory framework less efficient compared to its current implementation.

The parameters generated by both approaches were compared to the uniform coupler's fidelity in two types of simulations: probabilistic simulations and deterministic simulations.
In the probabilistic simulations, the mean fidelity and fidelity standard deviation is estimated for a range of different standard deviations used for the error distribution.
In the deterministic simulations, the fidelity was calculated for a range of deterministic errors between -3 and 3 standard deviations (width error between -20~nm and 20~nm).
In these simulations, both approaches presented parameters for segmented couplers, which were far more robust to systematic errors compared to the uniform coupler.

In the last section of the results, we show a clear connection between the reduction of the physical gate error using our optimized segmented couplers and the logical error rate PL in the quantum circuit. For a physical error  of 20 nm (roughly $\sim 5\%$ of the average waveguide width), both the uniform and segmented coupler were below the threshold of the quantum error correcting code, meaning both could potentially be corrected by the error correcting codes, but the uniform coupler required many more resources (qubits and quantum gates) to do so.
For a higher physical error , 60 nm  (roughly $\sim 15\%$ of the average waveguide width), the segmented coupler was still below the threshold, however, the uniform coupler was not, suggesting that it could not be corrected by the quantum error correcting code such as surface code.

As mentioned in the results, we uploaded a platform which allows testing the directional coupler's fidelity robustness increase in a QFT cirucit. In this platform the user can edit the geometric parameters and width error of the coupler, as well as the number of qubits used in the QFT, and observe how they affect the infidelity of the coupler.

\section{Conclusions \label{sec: Conclusions}}

To conclude, our composite approaches are shown to be far more robust to random systematic errors compared to the uniform coupler. We provide two approaches to construct these robust segmented gates: a perturbative approach and a non-perturbative approach, and demonstrated them in the photonic realm for the directional coupler realization of the gates. Specifically, we constructed robust designs against correlated Normally distributed width errors for the $X$, ${X}^{\frac{1}{2}}$, ${X}^{\frac{1}{3}}$, and $H$ gates.

The approaches shown in this paper were demonstrated with directional couplers, but the algorithms presented are by no means limited to optic-based quantum computation --- they are applicable to any quantum computing hardware, making both approaches (perturbative and non-perturbative) relevant in different quantum gate implementations as well.
Furthermore, while this paper concentrates on single-qubit gates and correlated errors, we expect the methodology to apply to multi-qubit gates or gates with partial error correlation between the segments, which are worthy topics for future studies.
Another important topic for further research is the evaluation of how segmented couplers can impact the success rate of success of state-of-the-art quantum error correcting codes, such as the surface code; this illustrates how the correction of systematic errors can increase the success rates of modern quantum algorithms. We believe that our segmented design for unitary gate operations and the design methods that were provided here could serve as fundamental elements and operations in many physical realizations of quantum information processing and quantum computing.

\section*{Acknowledgements}
Our work has been supported by the Israel Science
Foundation (ISF) and the Directorate for Defense Research and Development (DDR\&D) grant No. 3427/21.
M.G. has been further supported by the US-Israel Binational Science Foundation (BSF) Grant No. 2020072.
The work of Y.O. is supported in part by an ISF Center of Excellence.

\small
\bibliographystyle{unsrt.bst}
\bibliography{references}
\newpage

\appendix

\section{Perturbative Solutions for Fully Correlated Detuning Errors\label{appendix: Analytical Solutions for The Toy Model}}

\subsection{\texorpdfstring{$iX$}{Lg} gate in 3 segments}

A first-order solution for the $iX$ gate in 3 segments is given by:
\begin{subequations}
\begin{equation}
    \Omega_1=\Omega  , \;\; \Delta_1=\Delta  , \;\; t_1=\frac{\pi}{\sqrt{\Omega^2+\Delta^2}}  , 
\end{equation}
\begin{equation}
    \Omega_2=\frac{\Omega^2+\Delta^2}{2\Omega}  , \;\; \Delta_2=0  , \;\; t_2=\frac{2\pi\Omega}{\Omega^2+\Delta^2}  , 
\end{equation}
\begin{equation}
    \Omega_3=\Omega  , \;\; \Delta_3=-\Delta  , \;\; t_3=\frac{\pi}{\sqrt{\Omega^2+\Delta^2}}  , 
\end{equation}
\label{eq: iX gate in 3 segments}
\end{subequations}
\noindent%
where $\Omega>0$ and $\Delta$ are free parameters. Examples of this solution are fully presented in Table \ref{table: iX gate in 3 segments}, and their fidelities are shown in Fig.~\ref{fig: Analytical For Toy Model}(a). One can see clearly how the fidelity improves with the composite design.

\begin{table}[!h]
\centering
\resizebox{8.5cm}{!}{%
  \begin{tabular}{|c||c|c|c|}
 \hline
 \multicolumn{4}{|c|}{Composite $iX$ gate in 3 segments} \\
 \hline
 solution & {$\Omega_1,\Delta_1,t_1$} & {$\Omega_2,\Delta_2,t_2$} & {$\Omega_3,\Delta_3,t_3$}\\
 \hline
 \hline
 $\Delta_1=0.5\Omega$ & {1.00, 0.500, 2.81} & {0.625, 0, 5.03} & {1.00, -0.500, 2.81} \\
 $\Delta_1=0.75\Omega$ & {1.00, 0.750, 2.51} & {0.781, 0, 4.02} & {1.00, -0.750, 2.51} \\
 $\Delta_1=1\Omega$ & {1.00, 1.00, 2.22} & {1.00, 0, 3.14} & {1.00, -1.00, 2.22} \\
 $\Delta_1=1.1\Omega$ & {1.00, 1.1, 2.11326} & {1.105, 0, 2.84307} & {1.00, -1.1, 2.11326} \\
 $\Delta_1=1.2\Omega$ & {1.00, 1.2, 2.0112} & {1.22, 0, 2.57508} & {1.00, -1.2, 2.0112} \\
 \hline
  \end{tabular}}
  \caption{Examples of $iX$ gate generated by Eq.~(\ref{eq: iX gate in 3 segments}), such that the couplings and detunings are of the same order. The fidelity of these gates compared to the one-segments $iX$ gate are shown in Fig.~\ref{fig: Analytical For Toy Model}(a).}
  \label{table: iX gate in 3 segments}
\end{table}

\subsection{\texorpdfstring{${\left(iX\right)}^{\frac{1}{n}}$}{Lg} gate in 3 segments}

A first-order solution for the ${\left(iX\right)}^{\frac{1}{n}}$ gate in 3 segments is given by:
\begin{subequations}
\begin{equation}
    \Omega_1=\Omega  , \;\; \Delta_1=0  , \;\; t_1=\frac{\theta}{\Omega}  , 
\end{equation}
\begin{equation}
\begin{split}
    \Omega_2=\frac{\sin{\left(\theta+\frac{\pi}{2n} \right)}}{\sin{\left(\theta+\frac{\pi}{2n} \right)}-\sin{\left(\frac{\pi}{2n} \right)}} \Omega  , \;\; \Delta_2=0  , \\ \;\; t_2=\frac{2\left(2\pi m-\theta - \frac{\pi}{2n} \right)}{\Omega_2}  , 
\end{split}
\end{equation}
\begin{equation}
    \Omega_3=\Omega  , \;\; \Delta_3=0  , \;\; t_3=\frac{\theta}{\Omega}  , 
\end{equation}
\label{eq: (iX)^1/n gate in 3 segments}
\end{subequations}
where $\Omega>0$ and $\theta$ are free real parameters and $m$ is a free integer parameter (with the constraint $t_2>0$). Examples of this solution for $n=2,3$ are presented in Tables \ref{table: (iX)^1/2 gate in 3 segments},\ref{table: (iX)^1/3 gate in 3 segments},and their fidelities are shown in Figs. \ref{fig: Analytical For Toy Model}(b)+(c), where $m=1$ where chosen for all of them. One can see clearly how the fidelity improves with the composite design.

\begin{table}[!h]
\centering
\resizebox{8.5cm}{!}{%
  \begin{tabular}{|c||c|c|c|}
 \hline
 \multicolumn{4}{|c|}{Composite ${\left(iX\right)}^{\frac{1}{2}}$ gate in 3 segments} \\
 \hline
 solution & {$\Omega_1,\Delta_1,t_1$} & {$\Omega_2,\Delta_2,t_2$} & {$\Omega_3,\Delta_3,t_3$}\\
 \hline
 \hline
 $\theta=\frac{\pi}{2.2}$ & {1.00, 0, 1.428} & {8.56794, 0, 0.950004} & {1.00, 0, 1.428} \\
 $\theta=\frac{\pi}{2.4}$ & {1.00, 0, 1.309} & {5.44949, 0, 1.53731} & {1.00, 0, 1.309} \\
 $\theta=\frac{\pi}{2.6}$ & {1.00, 0, 1.2083} & {4.45279, 0, 1.92665} & {1.00, 0, 1.2083} \\
 $\theta=\frac{\pi}{2.8}$ & {1.00, 0, 1.122} & {3.98639, 0, 2.19537} & {1.00, 0, 1.122} \\
 $\theta=\frac{\pi}{3}$ & {1.00, 0, 1.05} & {3.73, 0, 2.39} & {1.00, 0, 1.05} \\
 $\theta=\frac{\pi}{4}$ & {1.00, 0, 0.785} & {3.41, 0, 2.76} & {1.00, 0, 0.785} \\
 $\theta=\frac{\pi}{5}$ & {1.00, 0, 0.628} & {3.52, 0, 2.77} & {1.00, 0, 0.628} \\
 \hline
  \end{tabular}}
    \caption{Examples of ${\left(iX\right)}^{\frac{1}{2}}$ gate generated by Eq. (\ref{eq: (iX)^1/n gate in 3 segments}) (with $n=2$ and choosing $m=1$), such that the couplings and detunings are of the same order. The fidelity of these gates compared to the one-segments ${\left(iX\right)}^{\frac{1}{2}}$ gate are shown in Fig.~\ref{fig: Analytical For Toy Model}(b).}
  \label{table: (iX)^1/2 gate in 3 segments}
\end{table}

\begin{table}[!h]
\centering
\resizebox{8.5cm}{!}{%
  \begin{tabular}{|c||c|c|c|}
 \hline
 \multicolumn{4}{|c|}{Composite ${\left(iX\right)}^{\frac{1}{3}}$ gate in 3 segments} \\
 \hline
 solution & {$\Omega_1,\Delta_1,t_1$} & {$\Omega_2,\Delta_2,t_2$} & {$\Omega_3,\Delta_3,t_3$}\\
 \hline
 \hline
 $\theta=\frac{\pi}{1.8}$ & {1.00, 0, 1.74533} & {2.87939, 0, 2.78827} & {1.00, 0, 1.74533} \\
 $\theta=\frac{\pi}{2.2}$ & {1.00, 0, 1.428} & {2.16722, 0, 3.99737} &{1.00, 0, 1.428} \\
 $\theta=\frac{\pi}{2.4}$ & {1.00, 0, 1.309} & {2.07313, 0, 4.29359} & {1.00, 0, 1.309} \\
 $\theta=\frac{\pi}{2.6}$ & {1.00, 0, 1.2083} & {2.02659, 0, 4.49157} & {1.00, 0, 1.2083} \\
 $\theta=\frac{\pi}{2.8}$ & {1.00, 0, 1.122} & {2.00562, 0, 4.62459} & {1.00, 0, 1.122} \\
 $\theta=\frac{\pi}{3}$ & {1.00, 0, 1.05} & {2.00, 0, 4.71} & {1.00, 0, 1.05} \\
 $\theta=\frac{\pi}{4}$ & {1.00, 0, 0.785} & {2.07, 0, 4.80} & {1.00, 0, 0.785} \\
 $\theta=\frac{\pi}{5}$ & {1.00, 0, 0.628} & {2.21, 0, 4.65} & {1.00, 0, 0.628} \\
 $\theta=\frac{\pi}{6}$ & {1.00, 0, 0.524} & {2.37, 0, 4.43} & {1.00, 0, 0.524} \\
 $\theta=\frac{\pi}{7}$ & {1.00, 0, 0.449} & {2.53, 0, 4.19} & {1.00, 0, 0.449} \\
 \hline
  \end{tabular}}
      \caption{Examples of ${\left(iX\right)}^{\frac{1}{3}}$ gate generated by Eq.~(\ref{eq: (iX)^1/n gate in 3 segments}) (with $n=3$ and choosing $m=1$) such that the couplings and detunings are of the same order. The fidelity of these gates compared to the one-segments ${\left(iX\right)}^{\frac{1}{3}}$ gate are shown in Fig.~\ref{fig: Analytical For Toy Model}(c).}
  \label{table: (iX)^1/3 gate in 3 segments}
\end{table}

\subsection{Other families of solutions for \texorpdfstring{${\left(iX\right)}^{\frac{1}{n}}$}{Lg}}
We found additional first-order solutions for ${\left(iX\right)}^{\frac{1}{n}}$:
\begin{enumerate}
    \item 
\begin{subequations}
\begin{equation}
    \Omega_1=\Omega  , \;\; \Delta_1=0  , \;\; t_1=\frac{\pi}{\Omega}  , 
\end{equation}
\begin{equation}
    \Omega_2=\frac{\Omega}{2}  , \;\; \Delta_2=0  , \;\; t_2=\frac{2\left(2\pi- \frac{\pi}{n} \right)}{\Omega}  , 
\end{equation}
\begin{equation}
    \Omega_3=\Omega  , \;\; \Delta_3=0  , \;\; t_3=\frac{\pi}{\Omega}  , 
\end{equation}
\label{eq: (iX)^1/n gate in 3 segments B}
\end{subequations}
where $\Omega>0$ is a free real parameter.
\item
\begin{subequations}
\begin{equation}
    \Omega_1=\Omega  , \;\; \Delta_1=\Delta  , \;\; t_1=\frac{2\pi}{\sqrt{\Omega^2+\Delta^2}}  , 
\end{equation}
\begin{equation}
\begin{split}
    \Omega_2= \frac{\left( \sqrt{\Omega^2+\Delta^2} \right)^{\frac{3}{2}}}{2\pi \Delta^2} \tan{\left( \frac{\pi}{2n} \right)}  , \;\; \Delta_2=0  , \\ \;\; t_2=\frac{2\left(2\pi- \frac{\pi}{2n} \right)}{\Omega_2}  , 
\end{split}
\end{equation}

\begin{equation}
    \Omega_3=\Omega  , \;\; \Delta_3=-\Delta  , \;\; t_3=\frac{2\pi}{\sqrt{\Omega^2+\Delta^2}}  , 
\end{equation}
\label{eq: (iX)^1/n gate in 3 segments C}
\end{subequations}
where $\Omega>0$ and $\Delta$ are free real parameters.

\item

\begin{subequations}

\begin{equation}
\begin{split}
    & \Omega_1=\Omega  , \;\; \Delta_1=\Delta  , \\ \;\; t_1=&\frac{2\left(m\pi-\arctan{\left( \sqrt{1+\frac{\Delta^2}{\Omega^2}} \tan{\left( \frac{\pi}{2n} \right)} \right)}\right)}{\sqrt{\Omega^2+\Delta^2}}  , 
\end{split}
\end{equation}

\begin{equation}
    \Omega_2= \frac{ \left(\Omega^2+\Delta^2\right) \sin{\left( \frac{\pi}{2n} \right)}}{2\Omega \sin{\left( \frac{\pi}{2n} \right)} -t_1 \Delta^2 \cos{\left( \frac{\pi}{2n} \right)}}  , \;\; \Delta_2=0  , \;\; t_2=\frac{\pi}{n \Omega_2}  , 
\end{equation}

\begin{equation}
\begin{split}
    & \Omega_3=\Omega  , \;\; \Delta_3=-\Delta  , \\ \;\; t_3=&\frac{2\left(m\pi-\arctan{\left( \sqrt{1+\frac{\Delta^2}{\Omega^2}} \tan{\left( \frac{\pi}{2n} \right)} \right)}\right)}{\sqrt{\Omega^2+\Delta^2}}  , 
\end{split}
\end{equation}
\label{eq: (iX)^1/n gate in 3 segments D}
\end{subequations}
where $\Omega>0$ and $\Delta$ are free real parameters, and $m$ is a free integer parameter (with the constraint $t_1>0$).

\end{enumerate}

\subsection{\texorpdfstring{$iX$}{Lg} gate in 4 segments with constant coupling}

A first-order solution for the $iX$ gate in 4 segments with equal couplings:
\begin{subequations}

\begin{equation}
    \Omega_1=\Omega  , \;\; \Delta_1=0  , \;\; t_1=\frac{4\arctan{\left( 1+\sqrt{2} \right)}}{\Omega}  , 
\end{equation}

\begin{equation}
    \Omega_2=\Omega  , \;\; \Delta_2=\xi \Omega  , \;\; t_2=\frac{2\pi}{\sqrt{\Omega^2+\Delta_2^2}}  , 
\end{equation}

\begin{equation}
    \Omega_3=\Omega  , \;\; \Delta_3=-\xi \Omega  , \;\; t_3=\frac{2\pi}{\sqrt{\Omega^2+\Delta_2^2}}  , 
\end{equation}

\begin{equation}
    \Omega_4=\Omega  , \;\; \Delta_4=0  , \;\; t_4=\frac{4\arctan{\left( 1+\sqrt{2} \right)}}{\Omega}  , 
\end{equation}
\label{eq: iX gate in 4 segments with same coupling}
\end{subequations}
where $\Omega$ is a free real parameter and $\xi$ is one of the two positive solutions of the equation $\left(2\pi \xi^2\right)^2=(1+\xi^2)^3$ (i.e., $\xi\approx0.461,6.033$). These two solutions are summarized in Table \ref{table: iX gate in 4 segments with same coupling}, and their fidelities are shown in Fig.~\ref{fig: Analytical For Toy Model}(d). One can see clearly how the fidelity improves with the composite design.

\begin{table}[tb]
\resizebox{8.5cm}{!}{%
  \begin{tabular}{|c||c|c|c|c|}
 \hline
 \multicolumn{5}{|c|}{Composite $iX$ gate in 4 segments with same coupling for each} \\
 \hline
 solution & {$\Omega_1,\Delta_1,t_1$} & {$\Omega_2,\Delta_2,t_2$} & {$\Omega_3,\Delta_3,t_3$} & {$\Omega_4,\Delta_4,t_4$}\\
 \hline
 \hline
 $\xi\approx0.46097$ & {1.00, 0, 4.71} & {1.00, 0.460966, 5.70612} & {1.00, -0.460966, 
   5.70612} & {1.00, 0, 4.71} \\
  $\xi\approx6.03285$ & {1.00, 0, 4.71} & {1.00, 6.03285, 1.02748} & {1.00, -6.03285, 
   1.02748} & {1.00, 0, 4.71} \\
 \hline
  \end{tabular}}
  \caption{The two 4-segments composite $iX$ gates given by Eq. (\ref{eq: iX gate in 4 segments with same coupling}). The fidelity of these gates compared to the one-segments $iX$ gate are shown in Fig.~\ref{fig: Analytical For Toy Model}(d).}
  \label{table: iX gate in 4 segments with same coupling}
\end{table}

\subsection{The fidelity of the composite gates \label{subAppendix: fidelity of analytical composite}}

\begin{figure}[tb]{}
         \centering
    \includegraphics[width=0.48\textwidth,page=6]{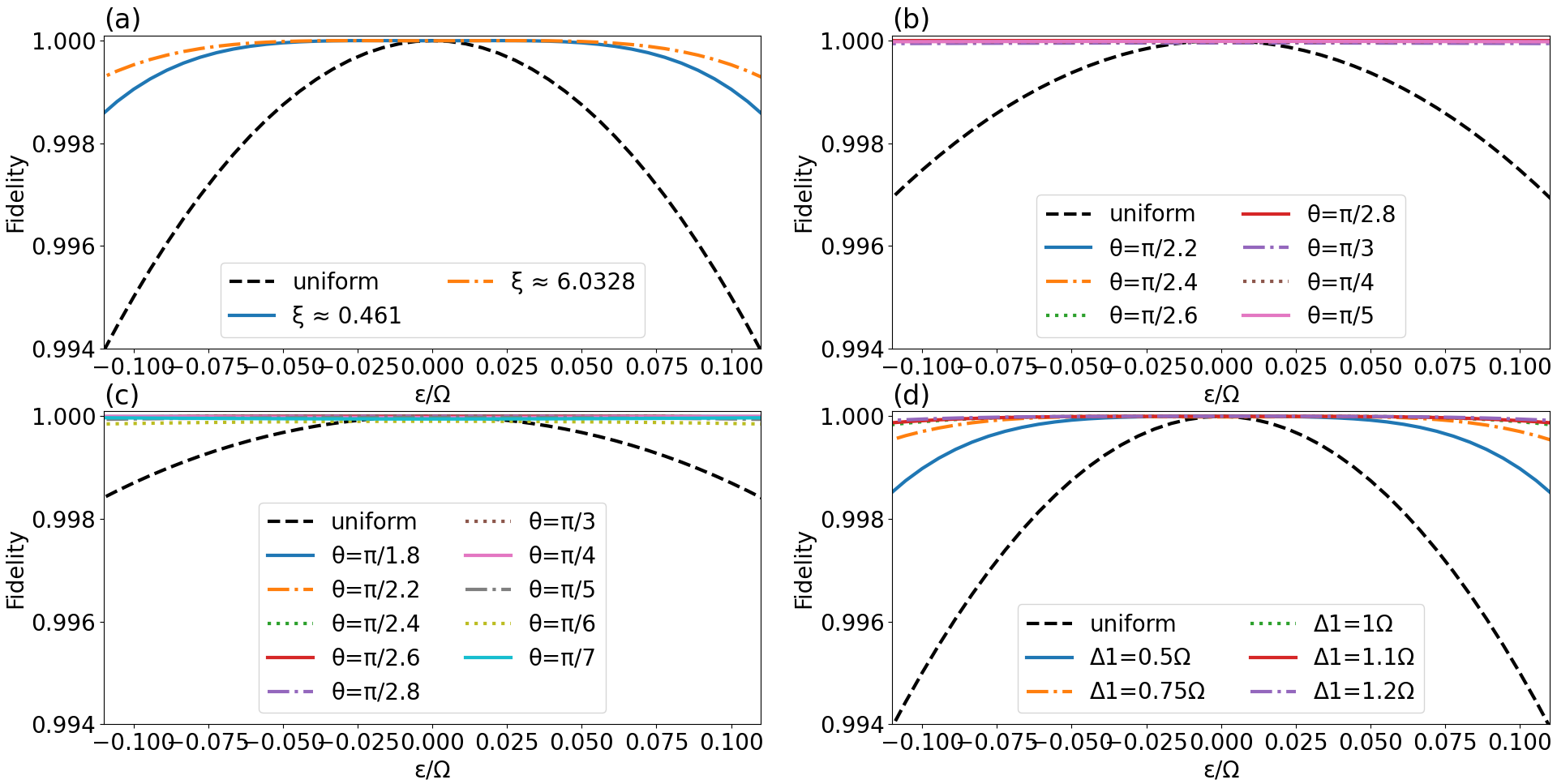}
         \caption{(a) The fidelity of examples of the composite $iX$ gate in 3 segments given by Eq.~(\ref{eq: iX gate in 3 segments}), compared to the one-segments case as a function of the detuning error. These parameters are given in Table~\ref{table: iX gate in 3 segments}. (b) The fidelity of examples of the composite ${\left(iX\right)}^{\frac{1}{2}}$ gate in 3 segments given by Eq. (\ref{eq: (iX)^1/n gate in 3 segments}), compared to the one-segments case. The parameters are given in Table \ref{table: (iX)^1/2 gate in 3 segments}. (c) The fidelity of examples of the composite ${\left(iX\right)}^{\frac{1}{3}}$ gate in 3 segments given by Eq. (\ref{eq: (iX)^1/n gate in 3 segments}), compared to the one-segments case. The parameters are given in Table \ref{table: (iX)^1/3 gate in 3 segments}. (d) The fidelity of the two 4-segment composite $iX$ gates given by Eq. (\ref{eq: iX gate in 4 segments with same coupling}), compared to the one-segments case.  The parameters are given in Table \ref{table: iX gate in 4 segments with same coupling}. We see the robustness of the segmented design in comparison to the uniform one.}
         \label{fig: Analytical For Toy Model}
     \end{figure}

In Fig.~\ref{fig: Analytical For Toy Model}, we present the results of the composite gates. We show the fidelity of our composite gates compared to regular uniform ones. The parameters of the segments were given in previous subsections of this section. Note that in these plots we consider the fidelity as a deterministic object and plot its values as a function of the values of the error.

\section{Non-Perturbative Solutions for Fully Correlated Detuning Errors\label{appendix: Non-Perturbative Solutions for The Toy Model}}
\begin{figure}[htbp]{}
         \centering
         \includegraphics[width=8.5cm]{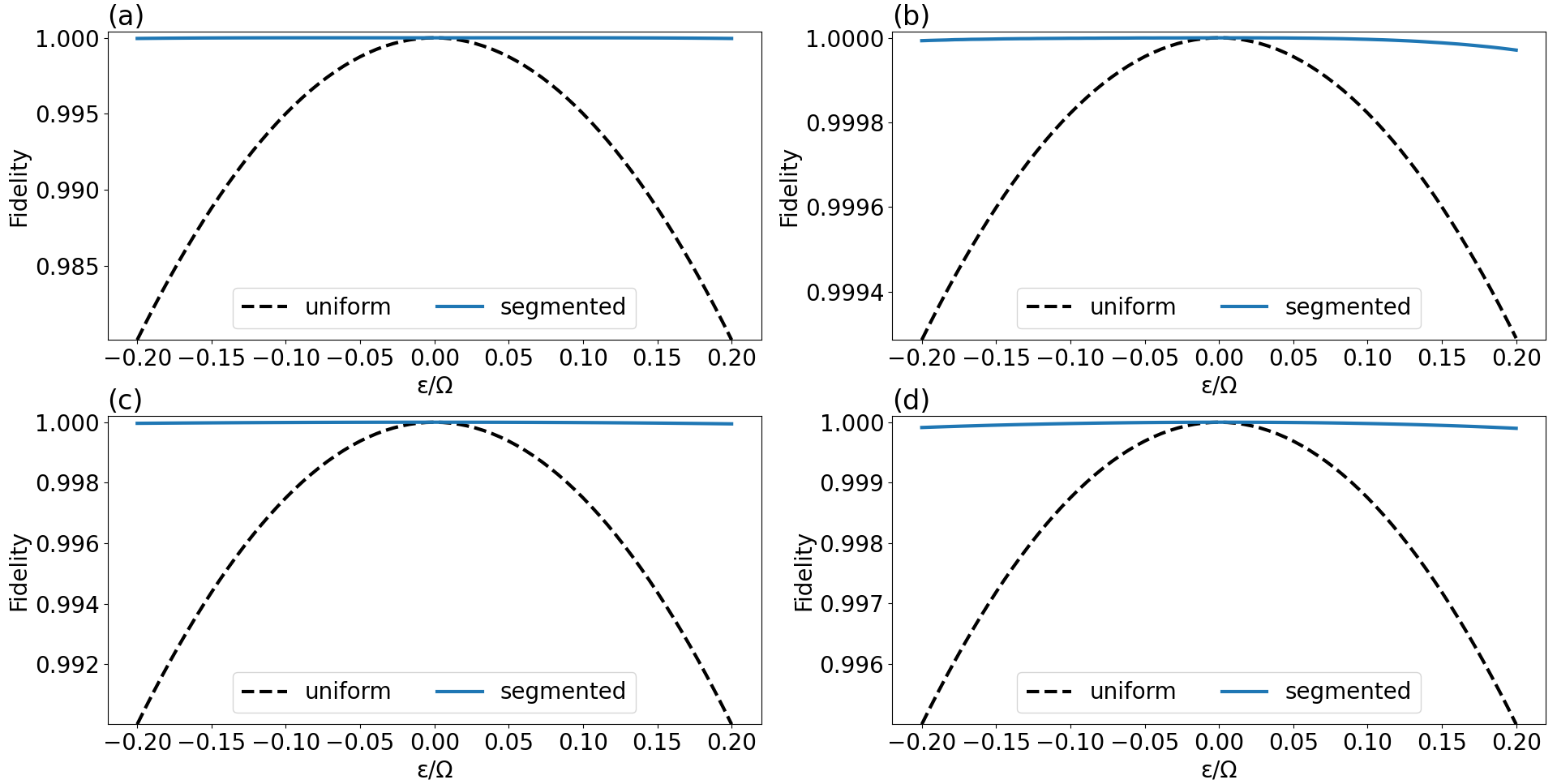}
         \caption{
         The fidelity of examples of the composite gates in 3 segments generated by the optimization algorithm compared to the one-segments case, as a function of the detuning error. In (a) the ideal gate is X, in (b) the ideal gate is the Hadamard gate, in (c) the ideal gate is $X^{\frac{1}{2}}$ and in (d) the ideal gate is $X^{\frac{1}{3}}$. All gates are calculated up to the global phase. These examples are given in Table \ref{table: Non-Perturbative examples For Toy Model}. We see the robustness of the segmented design in comparison to the uniform one.}
         \label{fig: Non Perturbative For Toy Model}
\end{figure}

Using the parameters optimization algorithm described in Section~\ref{appendix: numeric},  we generated optimization solutions for the following gates: $X$, $X^{\frac{1}{2}}$, $X^{\frac{1}{3}}$, $H$, resulting in the parameters given in Table~\ref{table: Non-Perturbative examples For Toy Model}.
Note that each of the gates constructed using these parameters is multiplied by a global phase.
The fidelity of the resulting gates is compared to the one-segments gates in Fig.~\ref{fig: Non Perturbative For Toy Model}.

\begin{table}[tb]
\centering
\resizebox{8.5cm}{!}{%
  \begin{tabular}{|c||c|c|c|}
 \hline
 \multicolumn{4}{|c|}{Composite gates in 3 segments} \\
 \hline
 Gate & {$\Omega_1,\Delta_1,t_1$} & {$\Omega_2,\Delta_2,t_2$} & {$\Omega_3,\Delta_3,t_3$}\\
 \hline
 \hline
 $X$ & {1.06, 1.784, 1.521} & {2.029, -0.005, 1.547} & {1.048, -1.776, 1.516} \\
 $X^{\frac{1}{2}}$ & {2.043, 0.2884, 2.0} & {5.763, -1.8525, 2.0} & {2.043, 0.2885, 2.0} \\
 $X^{\frac{1}{3}}$ & {3.629, 0.2737, 7.0} & {3.607, -0.4956, 7.0} & {3.6319, 0.259, 7.0} \\
 $H$ & {4.773, -0.978, 2.0} & {1.1855, 0.5415, 2.0} & {1.7075, -0.31135, 2.0} \\
 \hline
  \end{tabular}}
  \caption{Examples of non-perturbative solutions using the fully correlated detuning error model.}
  \label{table: Non-Perturbative examples For Toy Model}
\end{table}

\newpage

\section{Detailed explanation regarding the numeric approach methodology\label{appendix: numeric}}
In this section, we will describe in further details the optimization process of the numeric, non-perturbative method, and we will examplify the process using the detuning error model. As stated before, the non-perturbative approach generates optimized parameters by using a loss function which is composed of two subfunctions:
\begin{enumerate}
  \item Value range loss subfunction --- returns the sum:
  \begin{equation}
  \begin{split}
      \sum^{N-1}_{i=0} & \max(0,-t_i) +\max(0,\Omega_{\max} - \Omega_i) + \\ 
      & \max(0,\Omega_i - \Omega_{\min}) +
      \max(0,\Delta_{\max} - \Delta_i) + \\
      & \max(0, \Delta_i - \Delta_{\min} ) .
  \end{split}
  \end{equation}
  This subfunction ensures that the parameters we obtain are within their legal value range ($t_i$ values measure the length of the waveguides, which means they cannot be negative; the detuning and coupling parameters have a range of feasible values they can be in).
  \item  Robust fidelity loss --- returns the fidelity between the error-less matrix and a range of matrices created by using current parameters $\Omega_i,\Delta_i,t_i, i \in \{0,1,...,N-1\}$.
\end{enumerate}

This robust fidelity loss is calculated in the following way:
\begin{enumerate}
    \item Set vector $X$ to be a vector of n numbers evenly spaced between $-3\sigma$ to $3\sigma$; $X=[-3\sigma,-3\sigma+\frac{6\sigma}{n-1},-3\sigma+2\frac{6\sigma}{n-1},...,3\sigma-\frac{6\sigma}{n-1},3\sigma], \\$ 
    where $n$ is the number of error values used for the optimization (varies between optimization processes, in the range of 2,500 to 10,000). 
    
    \item Set vector Dist to be the given error distribution vector; for example, a Gaussian distributed vector is calculated as
$\mathrm{Dist} =[a_{-n/2+1},a_{-n/2+2},...,a_0,...,a_{n/2-2},a_{n/2-1}],$ where
$a_i = \frac{1}{\sigma \cdot \sqrt{2\pi}} \cdot e^{\frac{-x_i^2}{2\sigma^2}}$.

    \item Use the current parameters $\Omega_i,\Delta_i,t_i, i \in \{0,1,...,N-1\}$ and create n waveguide matrices, differing in the value of the error $\delta\Delta$; $\delta\Delta$ of the matrix $M_i$ is the element $x_i$ of the vector $X$: $$M_i=U_{3}(\Omega_0,\Delta_0+x_i,t_0,
          \Omega_1,\Delta_1+x_i,t_1,
          \Omega_2,\Delta_2+x_i,t_2).$$
          
    \item Calculate the fidelity loss of all these matrices and store them in vector $F$:
$$F_i = F_\mathrm{loss}(U_\mathrm{ideal},M_i),$$
where $F_\mathrm{loss}(U_\mathrm{ideal},U)= 1 - F_\mathrm{norm}(U_\mathrm{ideal},U)$
and $F_\mathrm{norm}(U_\mathrm{ideal},U)= |\mathrm{Tr}(U_\mathrm{ideal}^{\dagger}U)|/2 $
    \item  Return $F \cdot \mathrm{Dist}$ (scalar product between the vectors).
\end{enumerate}

Minimizing these subfunctions increases the overall fidelity robustness while keeping the parameters in their previously approved range. Generally, the optimizer used was the Adam optimizer with a learning rate of $10^{-3}$, but some gates were more delicate (for instance, $X^{0.5}$) and required a smaller learning rate. The optimization also worked well with a stochastic gradient descent optimizer.

\section{The parameters of directional couplers as a function of distance and widths\label{appendix: Fit}}

In order to estimate the detuning and coupling coefficients corresponding to the geometric parameters, we used Lumerical, a commercially available finite difference eigenmode solver.
We solved for the fields and effective mode indices $E(w), H(w), n(w)$ for different widths $w$.
With these solutions, we were able to approximate the dynamics parameters, using the coupled-mode theory perturbative approximation \cite{QuantumElectronicsYariv1989}:
\begin{subequations}
    \begin{equation}
    m_i \triangleq  \frac{\omega}{4}\int \int [\epsilon(x,y) - \epsilon^{(i)}(x,y)]
    \Big(\vec{E}_{\perp}(w_i)\Big)^2 dxdy ,
    \end{equation}
    \begin{equation}
        \Delta=\Delta\beta (w_1, w_2, g) \approx \frac{2\pi}{\lambda}(n_1 - n_2) + M_1 - M_2 ,
    \end{equation}
    \begin{equation}
    \begin{split}
        \Omega & = \kappa (w_1, w_2, g) \approx \\ & \frac{\omega}{4}\int \int [\epsilon(x,y) - \epsilon^{(2)}(x,y)]
        \vec{E}_{\perp}(w_1)\cdot \vec{E}_{\perp}(w_2) dxdy ,
\end{split}
    \end{equation}
    \label{eq: parameters from geometry}
\end{subequations}
where $\epsilon^{(i)}$ is defined as the permittivity distribution in space when only waveguide i exists. $m_1$ and $m_2$ represent small corrections to the propagation constants, $\beta_1$ and $\beta_2$, respectively, because of the presence of the second waveguide.

To be able to use this in the perturbative method or in gradient-based optimization algorithms, we performed this evaluation for a large number of geometries within our range of interest, and a multidimensional interpolation function was derived. The use of the coupled mode theory approximation enabled us to do so with a number of simulations that grows linearly with the number of different widths, and as a constant with respect to the number of gaps, instead of a number that grows as the number of widths squared times the number of gaps, as was needed for a more precise supermodes solution. The widths we took are between 300 and 400 nm, and the gaps we took are between 800 and 1200 nm. 

\begin{figure}[!h]
    \centering
    \includegraphics[width=6cm]{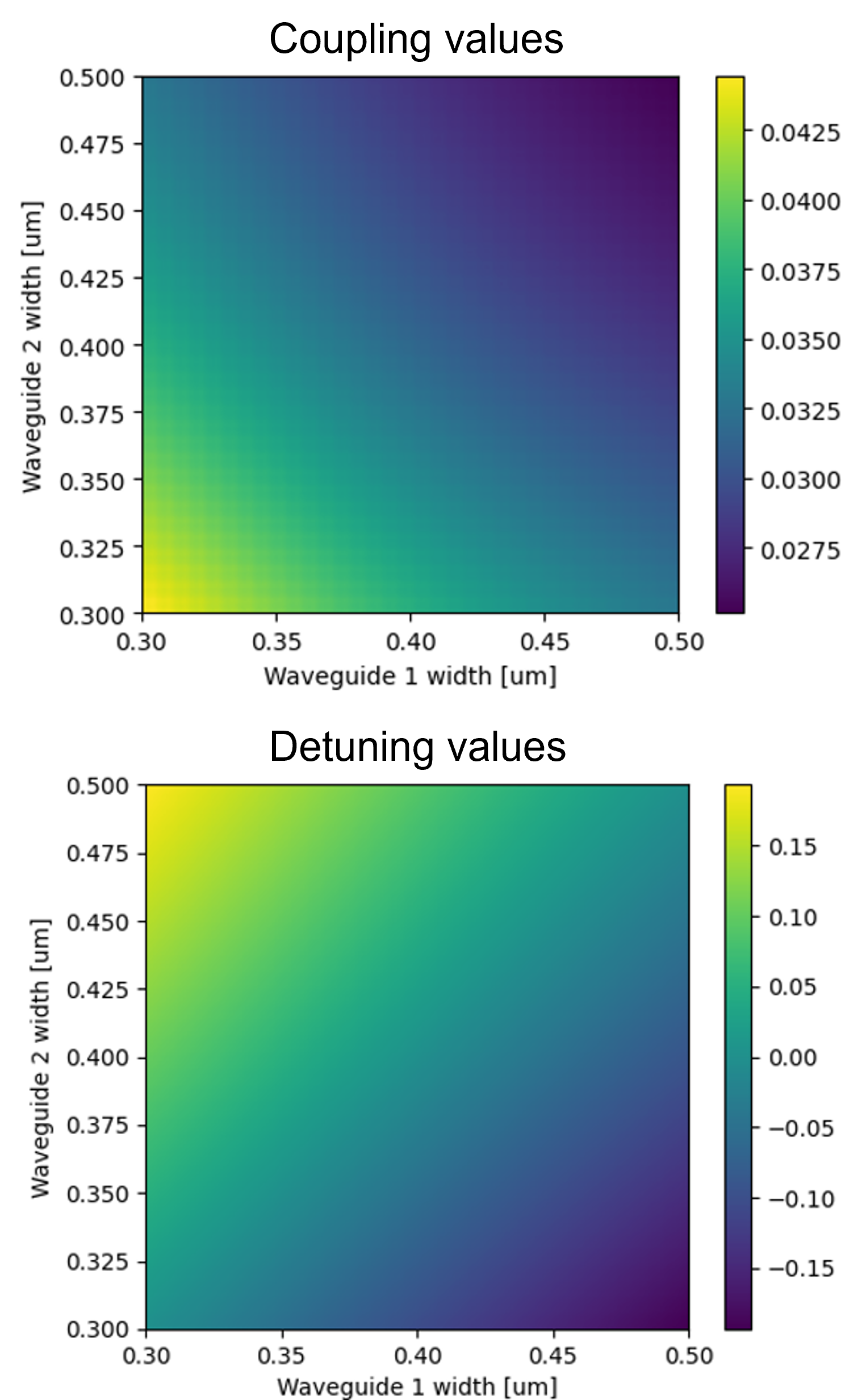}
    \caption{Grid of waveguide widths mapped to coupling and detuning coefficient values}
    \label{fig: coupling and detuning coefficient mapping}
\end{figure}

By calculating Eq.~(\ref{eq: parameters from geometry}) for different width values for the two waveguides, we obtain a grid of width values, which are correlated to a grid of detuning and coupling coefficients, as can be seen in figure \ref{fig: coupling and detuning coefficient mapping}.
After obtaining this grid, we used fitting algorithms in order to fit polynomial and exponential functions to the data. 

\begin{figure}[!h]
    \centering
    \includegraphics[width=6cm]{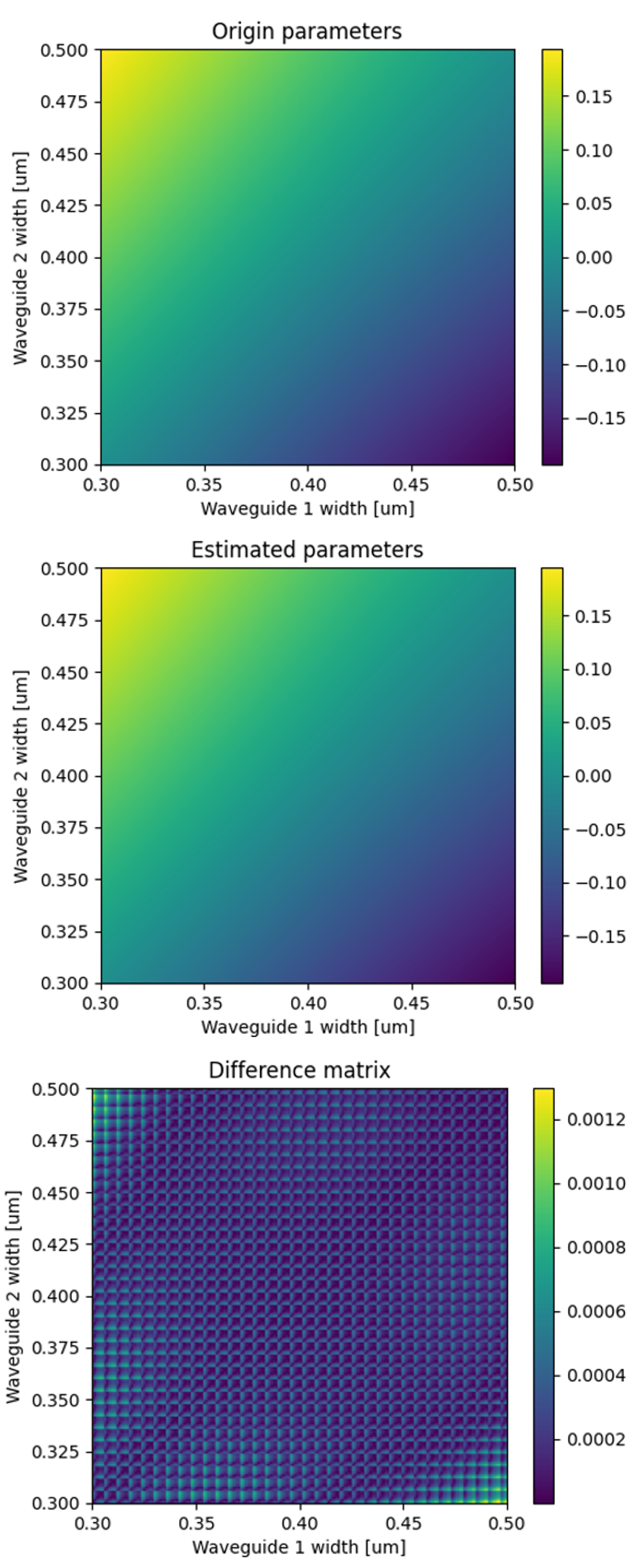}
    \caption{Comparison between the fit function estimation and the original detuning grid.}
    \label{fig: detuning coefficient estimating}
\end{figure}

For the detuning coefficient, a Taylor series in the form below was used (where $w_{i}$ is the width of waveguide $i$):
$$ \Delta = \sum_{i = 0}^{4} a_{i} \cdot  \text{w}_{1}^{i} + b_{i} \cdot  \text{w}_{2}^{i}\
$$
In figure \ref{fig: detuning coefficient estimating}, we can see how the approximate function behaves similarly to the values generated from the Lumerical simulations, where the average difference between the matrices is 0.00022, which translated to
$\sim 1$\% of the detuning coefficient.

\begin{figure}[tb]
    \centering
    \includegraphics[width=6cm]{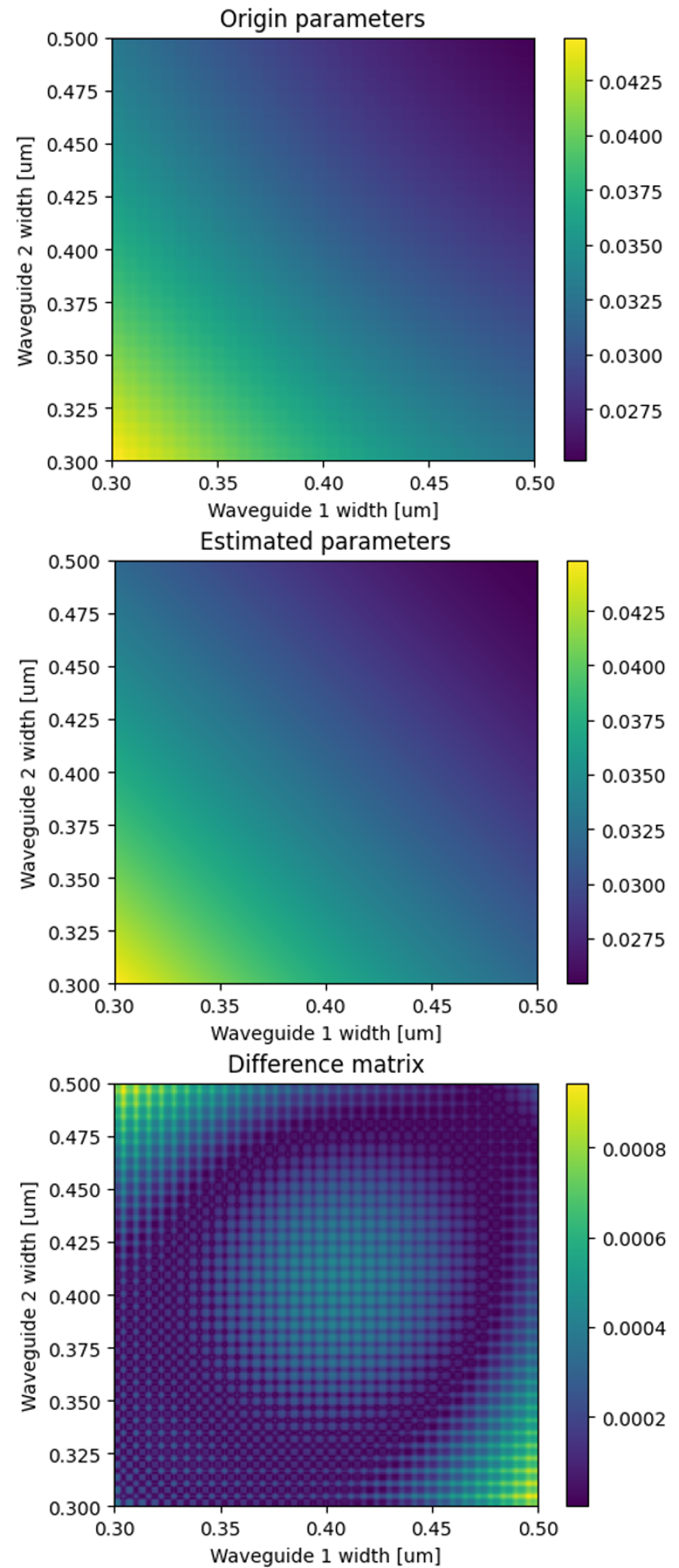}
    \caption{Comparison between the fit function estimation and the original coupling grid.}
    \label{fig: coupling coefficient estimating}
\end{figure}

For the Coupling coefficient, the following exponential function was used:
$$\Omega = a_{0} + 
a_{1} \cdot ( w_{1} + w_{2} ) \cdot  e^{a_{2} \cdot ( w_{1} + w_{2} )}
$$
In figure \ref{fig: coupling coefficient estimating}, we can see how the estimating function behaves similarly to the values generated from the CMT approximation, where the average difference between the matrices is 0.00017, which translated to \textasciitilde0.5\% of the coupling coefficient.

Lastly, we corrected the inaccuracies resulting from the coupled mode theory approximation as follows: We used the interpolation function to find the optimal composite design. We then calculated the expected rotation angle for each segment, $\theta = \Omega_g L$, where $\Omega_g = \sqrt{\Omega^2 + \Delta^2}$ is the generalized coupling coefficient. Then, in Lumerical, we solved for the generalized coupling coefficient of the selected segment using the more precise supermodes method~\cite{QuantumElectronicsYariv1989}. In this method, we solved for the modes of both waveguides together, unlike the coupled mode theory, where we solve for each mode separately and assume that the coupling is weak. The length of the segment is then determined from the desired rotation angle and the precise generalized coupling coefficient.

\newpage
\section{The geometrical parameters of the robust segmented gates \label{appendix: gate_parameters}}
Here we present selected solutions achieved from both approaches described in Sec.~\ref{sec: method for rho=1}. The parameters generated by the perturbative approach is presented in table \ref{table: Analytic robust gates against correlated errors in widths}, and parameters generated by the non-perturbative approach is presented in table \ref{table: Numeric robust gates against correlated errors in widths}. The gap between the two waveguides for all these solutions is 1.2$\mathrm{\mu m}$.

\newpage

\begin{table}[tb]
\centering
\resizebox{8.5cm}{!}{%
  \begin{tabular}{|c||c||c|c|c|}
 \hline
 \multicolumn{5}{|c|}{Composite gates in 3 segments based on the model of correlated errors of widths} \\
 \hline
 the gate & {${w_a}_{\mathrm{uni}},{w_b}_{\mathrm{uni}},z_{\mathrm{uni}}$} [$\mu$m] & {${w_a}_1,{w_b}_1,z_1$} [$\mu$m] & {${w_a}_2,{w_b}_2,z_2$} [$\mu$m] & {${w_a}_3,{w_b}_3,z_3$} [$\mu$m]\\
 \hline
 \hline
 $-iX$ &
 {0.450, 0.45,79.44} &
 { 0.4857, 0.4345,47.1117} &
 {0.4057, 0.4896, 40.5109} &
 {0.4857, 0.4345, 47.1117} \\
 ${\left(iX\right)}^{\frac{1}{2}}$ &
 {0.4, 0.4,118.972} &
 {0.426, 0.387,79.892} &
 {0.318, 0.499, 55.8067} &
 {0.426, 0.387, 79.892} \\
 $iH$ &
 {0.426, 0.460, 58.8037} &
 {0.379, 0.486,29.6481} &
 {0.5, 0.31, 53.75} &
 {0.379, 0.486, 29.6481} \\
 $\sqrt{\frac{1}{3}} I-i\sqrt{\frac{2}{3}}X$ &
 {0.450, 0.45,35.278} &
 {0.358, 0.457,21.890} &
 {0.485, 0.34, 28.0211} &
 {0.358, 0.457, 21.890} \\
 \hline
  \end{tabular}}
      \caption{Selected robust segmented gates achieved by the perturbative approach. These gates are robust against correlated errors in the widths.}
  \label{table: Analytic robust gates against correlated errors in widths}
\end{table}

\begin{table}[h!]
\centering
\resizebox{8.5cm}{!}{%
  \begin{tabular}{|c||c||c|c|c|}
 \hline
 \multicolumn{5}{|c|}{Composite gates in 3 segments based on the model of correlated errors of widths} \\
 \hline
 the gate & {${w_a}_{\mathrm{uni}},{w_b}_{\mathrm{uni}},z_{\mathrm{uni}}$} [$\mu$m] & {${w_a}_1,{w_b}_1,z_1$} [$\mu$m] & {${w_a}_2,{w_b}_2,z_2$} [$\mu$m] & {${w_a}_3,{w_b}_3,z_3$} [$\mu$m]\\
 \hline
 \hline
 $-iX$ &
 {0.450, 0.45,79.44} &
 {0.375, 0.425, 49.254} &
 {0.429, 0.363, 52.608} &
 {0.391, 0.45, 46.63} \\
 ${\left(iX\right)}^{\frac{1}{2}}$ &
 {0.4, 0.4,20.0872} &
 {0.48, 0.326, 15.28} &
 {0.32, 0.478, 28.402} &
 {0.48, 0.324, 15.18} \\
 $iH$ &
 {0.426, 0.460, 58.8037} &
 {0.430, 0.452, 70.29} &
 {0.422, 0.325, 33.522} &
 {0.430, 0.452, 70.328} \\
 $\sqrt{\frac{1}{3}} I-i\sqrt{\frac{2}{3}}X$ &
 {0.450, 0.45,35.278} &
 {0.351, 0.46, 20.468} &
 {0.459, 0.34, 34.078} &
 {0.349, 0.46, 20.261} \\
 \hline
  \end{tabular}}
      \caption{Selected robust segmented gates achieved by the non-perturbative approach. These gates are robust against correlated errors in the widths.}
  \label{table: Numeric robust gates against correlated errors in widths}
\end{table}

\section{The fidelity of composite gates using different error distributions}
\begin{figure}[tb]
    \centering
    \includegraphics[width=9cm]{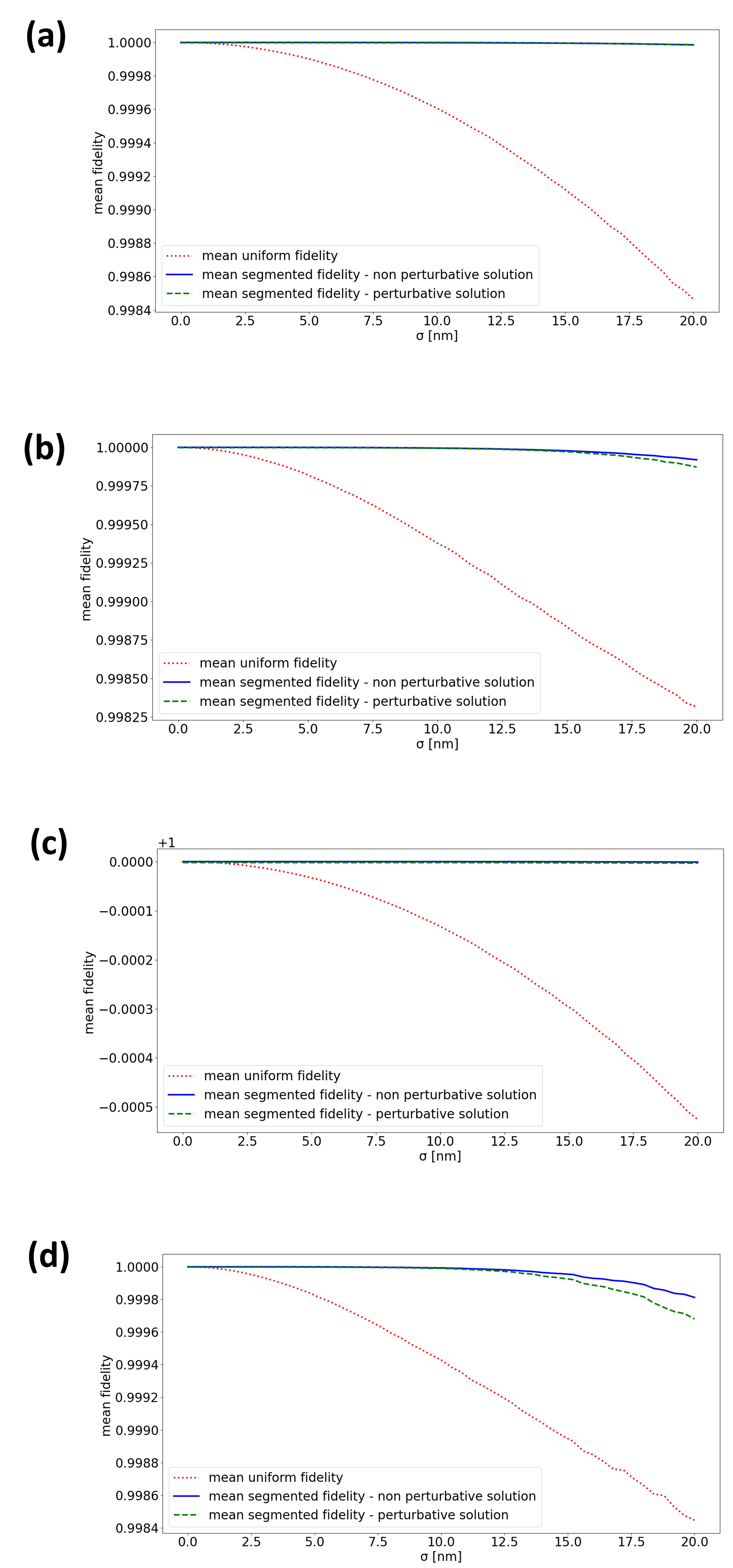}
    \caption{The mean fidelity of the pertubative and non-pertubative composite gates compared to uniform gates, with full error correlation in the width, as a function of the error standard deviation $\sigma$, using different error distributions, where the ideal gate is $X$. In (a) the error distribution is Gaussian, in (b) the error distribution is Poisson (where $\lambda=1$ and the result is multiplied by the value of $\sigma$), in (c) the error distribution is Uniform (uniform distribution between -3 $\sigma$ and 3 $\sigma$) and in (d) the error distribution is exponential.}
    \label{fig: different distributions}
\end{figure}

In order to ensure that the results presented in the article aren't only relevant to Gaussian error distribution, we added simulations of the fidelity using various additional error distributions, as can be seen in the Fig ~\ref{fig: different distributions}.

\end{document}